\documentclass[twocolumn]{aastex63}
\usepackage{graphicx}
\usepackage{float}
\usepackage{lipsum}
\usepackage{amsfonts,amsmath,amssymb}
\usepackage{natbib}
\usepackage{url}
\usepackage{xcolor}
\usepackage{hyperref}
\usepackage[numbered]{bookmark}
\usepackage{calculator}
\usepackage{placeins} % provides \FloatBarrier
\let\splitbox\relax
\usepackage{adjustbox} % Can set percentages when doing Clip in \adjincludegraphics. Note the uppercase Clip.
\usepackage{enumitem} % begin{itemize}[leftmargin=-3mm]
\usepackage{xifthen}
\usepackage{chngcntr} % for Figure numbering inside Appendix

\usepackage{bm}

\usepackage{tabularx}

\usepackage{adjustbox}

\hypersetup{
    colorlinks,
    linkcolor={magenta!80!white},
    citecolor={blue!80!black},
    urlcolor={magenta!80!black},
}

\usepackage{listings} % also contains \lstinline
\usepackage{color}
\definecolor{lightgray}{gray}{0.9}

\usepackage{lmodern}% for \textsl{\textsc{SuperDeblend}}
\usepackage{slantsc}% for \textsl{\textsc{SuperDeblend}}

\usepackage{multirow}

\newcommand{\TODO}[2][]{%
\ifthenelse{\isempty{#1}}%
{\textbf{\textcolor{red}{TODO: #2}}}%
{\textbf{\textcolor{red}{TODO (#1) #2}}}%
}

\def\gtsima{$\; \buildrel > \over \sim \;$}
\def\ltsima{$\; \buildrel < \over \sim \;$}
\def\prosima{$\; \buildrel \propto \over \sim \;$}
\def\gsim{\lower.5ex\hbox{\gtsima}}
\def\lsim{\lower.5ex\hbox{\ltsima}}
\def\simgt{\lower.5ex\hbox{\gtsima}}
\def\simlt{\lower.5ex\hbox{\ltsima}}
\def\simpr{\lower.5ex\hbox{\prosima}}

\def\beq{\begin{equation}}
\def\eeq{\end{equation}}
\def\8mu{8\,$\mu{\rm m}$}
\def\16mu{16\,$\mu{\rm m}$}
\def\24mu{24\,$\mu{\rm m}$}
\def\70mu{70\,$\mu{\rm m}$}

\def\alphaCO{\alpha_{\mathrm{CO}}}

\def\Tkin{T_{\mathrm{kin}}}

\def\nH2{n_{\mathrm{H}_2}}
\def\nHtwo{n_{\mathrm{H}_2}}
\def\NH2_{N_{\mathrm{H}_2}}

\def\lgNH2_{\mathrm{log}_{10}\,(N_{\mathrm{H}_2} / \mathrm{cm^{-2}})}

\def\lgnHtwo{\mathrm{log}\,n_{\mathrm{H_2}}}

\def\HIrom{\mathrm{H}\textnormal{\textup{\uppercase\expandafter{\romannumeral 1}}}}
\def\HIIrom{\mathrm{H}\textnormal{\textup{\uppercase\expandafter{\romannumeral 2}}}}

\def\Msun{\mathrm{M}_{\odot}}

\def\Kkmspc2{\mathrm{K\,km\,s^{-1}\,pc^{2}}}

\def\R52{R_{52}}

\def\fH2_{f_{\mathrm{H_2}}}
\def\MH2_{M_{\mathrm{H_2}}}
\def\XCICO{[\mathrm{C}{\text{\sc{i}}/\mathrm{CO}]}}

\def\Kkms_{\mathrm{K\,km/s}}

\def\michi2{\textsc{MiChi2}}

\def\um{\ifmmode{\mu\mathrm{m}}\else{$\mu\mathrm{m}$}\fi}

\newcommand{\J}[2][]{%
	\ifthenelse{\isempty{#1}}%
	{\begingroup%
		\def\JlowerN{}%
		\ADD{#2}{-1}{\JlowerN}%
		J=#2{\to}\JlowerN%
		\endgroup%
	}%
	{J=#1{\to}#2}%
}

\newcommand{\rom}[1]{{{\uppercase\expandafter{\romannumeral #1}}}}
\newcommand{\romup}[1]{{\textup{\uppercase\expandafter{\romannumeral #1}}}}

\newcommand{\incode}[1]{{\raggedright\lstinline|#1|}}
\newcommand{\incodep}[1]{{\raggedright\lstinline|"#1"|}}

\newcommand{\neutralcarbon}{\ifmmode \text{C\textsc{i}} \else {\sc C\,i}\fi}
\newcommand{\ionizedcarbon}{\ifmmode \text{C\textsc{ii}} \else {\sc C\,ii}\fi}
\newcommand{\neutralhydrogen}{\ifmmode \text{H\textsc{i}} \else {\sc H\,i}\fi}
\newcommand{\ionizedhydrogen}{\ifmmode \text{H\textsc{ii}} \else {\sc H\,ii}\fi}
\newcommand{\carbonmonoxide}{\ifmmode \text{\textsc{CO}} \else {\sc CO}\fi}

\def\CI{\neutralcarbon}

\def\alphaCO{\relax\ifmmode%
    \alpha_\mathrm{CO}
    \else{$\alpha_\mathrm{CO}$}\fi%
    }

\newcommand{\DefineRemark}[2]{%
\expandafter\newcommand\csname rmk-#1\endcsname{#2}%
}
\newcommand{\Remark}[1]{\csname rmk-#1\endcsname}
\DefineRemark{CI3P13P0}{[\mathrm{C}\,{\text{\sc{i}}}]\;(^{3}P_{1}\textnormal{--}^{3}P_{0})}
\DefineRemark{CI}{\mathrm{C}\,{\text{\sc{i}}}}
\DefineRemark{CI10}{[\mathrm{C}\,{\text{\sc{i}}}]\,(1\textnormal{--}0)}
\DefineRemark{CI21}{[\mathrm{C}\,{\text{\sc{i}}}]\,(2\textnormal{--}1)}
\DefineRemark{CO}{\mathrm{CO}}
\DefineRemark{CO10}{\mathrm{CO}\,(1\textnormal{--}0)}
\DefineRemark{CO21}{\mathrm{CO}\,(2\textnormal{--}1)}
\DefineRemark{CO32}{\mathrm{CO}\,(3\textnormal{--}2)}
\DefineRemark{CO43}{\mathrm{CO}\,(4\textnormal{--}3)}
\DefineRemark{COJ10}{\mathrm{CO}\,(J=1 \to 0)}
\DefineRemark{COJ21}{\mathrm{CO}\,(J=2 \to 1)}
\DefineRemark{COJ32}{\mathrm{CO}\,(J=3 \to 2)}
\DefineRemark{COJ43}{\mathrm{CO}\,(J=4 \to 3)}

\DefineRemark{RCI10CO10}{%
{R}_{\mathrm{C{\text{\sc{i}}}_{10}}/\mathrm{CO_{10}}}%
}
\DefineRemark{RCI10CO21}{%
{R}_{\mathrm{C{\text{\sc{i}}}_{10}}/\mathrm{CO_{21}}}%
}
\DefineRemark{RCI10CO32}{%
{R}_{\mathrm{C{\text{\sc{i}}}_{10}}/\mathrm{CO_{32}}}%
}
\DefineRemark{RCI10CO43}{%
{R}_{\mathrm{C{\text{\sc{i}}}_{10}}/\mathrm{CO_{43}}}%
}

\DefineRemark{LPrimeCO21}{L^{\prime}_{\Remark{CO21}}}
\DefineRemark{LPrimeCI10}{L^{\prime}_{\Remark{CI10}}}
\DefineRemark{ICO10}{I_{\Remark{CO10}}}
\DefineRemark{ICO21}{I_{\Remark{CO21}}}
\DefineRemark{ICO32}{I_{\Remark{CO32}}}
\DefineRemark{ICO43}{I_{\Remark{CO43}}}
\DefineRemark{ICI10}{I_{\Remark{CI10}}}

\DefineRemark{kms}{\mathrm{km\,s^{-1}}}
\DefineRemark{Kkms}{\mathrm{K\,km\,s^{-1}}}
\DefineRemark{Kkmspc2}{\mathrm{K\,km\,s^{-1}\,pc^{2}}}
\DefineRemark{MJysr}{\mathrm{MJy\,sr^{-1}}}

\DefineRemark{R42}{\ifmmode{R_{42}}\else{$R_{42}$}\fi}
\DefineRemark{R41}{\ifmmode{R_{41}}\else{$R_{41}$}\fi}
\DefineRemark{R31}{\ifmmode{R_{31}}\else{$R_{31}$}\fi}
\DefineRemark{R21}{\ifmmode{R_{21}}\else{$R_{21}$}\fi}
\DefineRemark{R02}{\ifmmode{R_{\rm C\text{\sc{i}}CO}}\else{$R_{\rm C\text{\sc{i}}CO}$}\fi}
\DefineRemark{RCICO}{\ifmmode{R_{\mathrm{C}\text{\sc{i}}\mathrm{CO}}}\else{$R_{\mathrm{C}\text{\sc{i}}\mathrm{CO}}$}\fi}

\renewcommand{\figurename}{Fig.}

\accepted{December 7, 2022}
\submitjournal{ApJL}
\shorttitle{Stellar Feedback, Molecular Excitation and Dissociation in NGC~1365}
\shortauthors{Liu et al.}

\begin{document}

\title{PHANGS-JWST First Results: Stellar Feedback-Driven Excitation and Dissociation of Molecular Gas in the Starburst Ring of NGC 1365?
}

\correspondingauthor{Daizhong Liu}
\email{dzliu@mpe.mpg.de, astro.dzliu@gmail.com}

%----- WRITING, ANALYSIS -----%

\author[0000-0001-9773-7479]{Daizhong Liu}
\affiliation{Max-Planck-Institut f\"ur Extraterrestrische Physik (MPE), Giessenbachstr. 1, D-85748 Garching, Germany}

\author[0000-0002-3933-7677]{Eva Schinnerer}
\affiliation{Max-Planck-Institut f\"ur Astronomie, K\"onigstuhl 17, D-69117 Heidelberg, Germany}

\author[0000-0001-5301-1326]{Yixian Cao}
\affiliation{Max-Planck-Institut f\"ur Extraterrestrische Physik (MPE), Giessenbachstr. 1, D-85748 Garching, Germany}

\author[0000-0002-2545-1700]{Adam Leroy}
\affiliation{Department of Astronomy, The Ohio State University, 140 West 18th Ave, Columbus, OH 43210, USA}

\author[0000-0003-1242-505X]{Antonio Usero}
\affiliation{Observatorio Astron\'{o}mico Nacional (IGN), C/Alfonso XII, 3, E-28014 Madrid, Spain}

\author[0000-0002-5204-2259]{Erik Rosolowsky}
\affiliation{Department of Physics, University of Alberta, Edmonton, AB T6G 2E1, Canada}

\author[0000-0002-6155-7166]{Eric Emsellem}
\affiliation{European Southern Observatory, Karl-Schwarzschild-Stra{\ss}e 2, 85748 Garching, Germany}
\affiliation{Univ Lyon, Univ Lyon1, ENS de Lyon, CNRS, Centre de Recherche Astrophysique de Lyon UMR5574, F-69230 Saint-Genis-Laval France}

%----- COMMENTING (SORTED BY IMPORTANCE/NUMBER OF COMMENTS IN OVERLEAF) -----%

\author[0000-0002-8804-0212]{J.~M.~Diederik~Kruijssen}
\affiliation{Cosmic Origins Of Life (COOL) Research DAO, coolresearch.io}

\author[0000-0002-5635-5180]{M\'elanie Chevance}
\affiliation{Universit\"at Heidelberg, Zentrum f\"ur Astronomie, Institut f\"ur Theoretische Astrophysik, Albert-Ueberle-Str 2, D-69120 Heidelberg, Germany}
\affiliation{Cosmic Origins Of Life (COOL) Research DAO, coolresearch.io}

\author[0000-0001-6708-1317]{Simon C.\ O.\ Glover}
\affiliation{Universit\"at Heidelberg, Zentrum f\"ur Astronomie, Institut f\"ur Theoretische Astrophysik, Albert-Ueberle-Str 2, D-69120 Heidelberg, Germany}

\author[0000-0001-6113-6241]{Mattia C.\ Sormani}
\affiliation{Universit\"at Heidelberg, Zentrum f\"ur Astronomie, Institut f\"ur Theoretische Astrophysik, Albert-Ueberle-Str 2, D-69120 Heidelberg, Germany}

\author[0000-0002-5480-5686]{Alberto D. Bolatto}
\affiliation{Department of Astronomy and Joint Space-Science Institute, University of Maryland, College Park, MD 20742, USA}

\author[0000-0003-0378-4667]{Jiayi~Sun}
\affiliation{Department of Physics and Astronomy, McMaster University, 1280 Main Street West, Hamilton, ON L8S 4M1, Canada}
\affiliation{Canadian Institute for Theoretical Astrophysics (CITA), University of Toronto, 60 St George Street, Toronto, ON M5S 3H8, Canada}

\author[0000-0002-9333-387X]{Sophia K. Stuber}
\affiliation{Max-Planck-Institut f\"ur Astronomie, K\"onigstuhl 17, D-69117 Heidelberg, Germany}

\author[0000-0003-4209-1599]{Yu-Hsuan Teng}
\affiliation{Center for Astrophysics and Space Sciences, Department of Physics, University of California, San Diego, 9500 Gilman Drive, La Jolla, CA 92093, USA}

\author[0000-0003-0166-9745]{Frank Bigiel}
\affiliation{Argelander-Institut f\"{u}r Astronomie, Universit\"{a}t Bonn, Auf dem H\"{u}gel 71, 53121 Bonn, Germany}

\author[0000-0003-0583-7363]{Ivana Be\v{s}li\'c}
\affiliation{Argelander-Institut f\"{u}r Astronomie, Universit\"{a}t Bonn, Auf dem H\"{u}gel 71, 53121 Bonn, Germany}

\author[0000-0002-3247-5321]{Kathryn~Grasha}
\affiliation{Research School of Astronomy and Astrophysics, Australian National University, Canberra, ACT 2611, Australia}
\affiliation{ARC Centre of Excellence for All Sky Astrophysics in 3 Dimensions (ASTRO 3D), Australia}

\author[0000-0001-9656-7682]{Jonathan~D.~Henshaw}
\affiliation{Astrophysics Research Institute, Liverpool John Moores University, 146 Brownlow Hill, Liverpool L3 5RF, UK}
\affiliation{Max-Planck-Institut f\"ur Astronomie, K\"onigstuhl 17, D-69117 Heidelberg, Germany}

\author[0000-0003-0410-4504]{Ashley.~T.~Barnes}
\affiliation{Argelander-Institut f\"{u}r Astronomie, Universit\"{a}t Bonn, Auf dem H\"{u}gel 71, 53121 Bonn, Germany}

\author[0000-0002-8760-6157]{Jakob~S.~den Brok}
\affiliation{Argelander-Institut f\"{u}r Astronomie, Universit\"{a}t Bonn, Auf dem H\"{u}gel 71, 53121 Bonn, Germany}

\author[0000-0002-2501-9328]{Toshiki Saito}
\affiliation{National Astronomical Observatory of Japan, 2-21-1 Osawa, Mitaka, Tokyo, 181-8588, Japan}

\author[0000-0002-5782-9093]{Daniel~A.~Dale}
\affiliation{Department of Physics and Astronomy, University of Wyoming, Laramie, WY 82071, USA}

\author[0000-0002-7365-5791]{Elizabeth~J.~Watkins}
\affiliation{Astronomisches Rechen-Institut, Zentrum f\"{u}r Astronomie der Universit\"{a}t Heidelberg, M\"{o}nchhofstra\ss e 12-14, 69120 Heidelberg, Germany}

\author[0000-0002-1370-6964]{Hsi-An Pan}
\affiliation{Department of Physics, Tamkang University, No.151, Yingzhuan Road, Tamsui District, New Taipei City 251301, Taiwan}

\author[0000-0002-0560-3172]{Ralf S.\ Klessen}
\affiliation{Universit\"at Heidelberg, Zentrum f\"ur Astronomie, Institut f\"ur Theoretische Astrophysik, Albert-Ueberle-Str 2, D-69120 Heidelberg, Germany}
\affiliation{Universit\"at Heidelberg, Interdisziplin\"ares Zentrum f\"ur Wissenschaftliches Rechnen, Im Neuenheimer Feld 205, D-69120 Heidelberg, Germany}

%----- DATA, OBERVATION -----%

\author[0000-0002-5259-2314]{Gagandeep~S. Anand}
\affiliation{Space Telescope Science Institute, 3700 San Martin Drive, Baltimore, MD 21218, USA}

\author[0000-0003-1943-723X]{Sinan Deger}
\affiliation{The Oskar Klein Centre for Cosmoparticle Physics, Department of Physics, Stockholm University, AlbaNova, Stockholm, SE-106 91, Sweden}

\author[0000-0002-4755-118X]{Oleg V. Egorov}
\affiliation{Astronomisches Rechen-Institut, Zentrum f\"{u}r Astronomie der Universit\"{a}t Heidelberg, M\"{o}nchhofstra\ss e 12-14, 69120 Heidelberg, Germany}

\author[0000-0001-5310-467X]{Christopher~M. Faesi}
\affiliation{University of Massachusetts-Amherst, 710 North Pleasant Street, Amherst, MA 01003, USA}

\author[0000-0002-8806-6308]{Hamid Hassani}
\affiliation{Department of Physics, University of Alberta, Edmonton, Alberta, T6G 2E1, Canada}

\author[0000-0003-3917-6460]{Kirsten L. Larson}
\affiliation{AURA for the European Space Agency (ESA), Space Telescope Science Institute, 3700 San Martin Drive, Baltimore, MD 21218, USA}

\author[0000-0002-2278-9407]{Janice~C. Lee}
\affiliation{Gemini Observatory/NSF's NOIRLab, 950 N. Cherry Avenue, Tucson, AZ 85719, USA; Caltech/IPAC, California Institute of Technology, Pasadena, CA 91125, USA}

\author[0000-0002-1790-3148]{Laura A. Lopez}
\affiliation{Department of Astronomy, The Ohio State University, 140 West 18th Ave, Columbus, OH 43210, USA}
\affiliation{Center for Cosmology and Astroparticle Physics, 191 West Woodruff Avenue, Columbus, OH 43210, USA}
\affiliation{Flatiron Institute, Center for Computational Astrophysics, NY 10010, USA}

\author[0000-0003-3061-6546]{J\'{e}r\^{o}me Pety}
\affiliation{IRAM, 300 rue de la Piscine, 38400 Saint Martin d'H\`eres, France}
\affiliation{LERMA, Observatoire de Paris, PSL Research University, CNRS, Sorbonne Universit\'es, 75014 Paris}

\author[0000-0002-4378-8534]{Karin Sandstrom}
\affiliation{Center for Astrophysics and Space Sciences, Department of Physics, University of California, San Diego, 9500 Gilman Drive, La Jolla, CA 92093, USA}

\author[0000-0002-8528-7340]{David~A. Thilker}
\affiliation{Department of Physics and Astronomy, The Johns Hopkins University, Baltimore, MD 21218, USA}

\author[0000-0002-3784-7032]{Bradley~C. Whitmore}
\affiliation{Space Telescope Science Institute, 3700 San Martin Drive, Baltimore, MD 21218, USA}

\author[0000-0002-0012-2142]{Thomas G. Williams}
\affiliation{Sub-department of Astrophysics, Department of Physics, University of Oxford, Keble Road, Oxford OX1 3RH, UK}
\affiliation{Max-Planck-Institut f\"ur Astronomie, K\"onigstuhl 17, D-69117 Heidelberg, Germany}

%\author{PHANGS COLLABORATORS, PLEASE JOIN! (ORDER TBD)}
%\author{(ORDER TBD)}
%\author{\textcolor{red}{(PLEASE CHECK NAME SPELLING, ORCID AND AFFILIATIONS ON THE LAST PAGE!)}} % TODO: comment out in final version

% Co-Is in 2019.1.01635.S : Frank Bigiel, Melanie Chevance, Christopher Faesi, Simon Glover, Brent Groves, Cinthya Herrera, I-TIng Ho, Annie Hughes, Alex Hygate, Kathryn Kreckel, Diederik Kruijssen, Adam Leroy, Eve Ostriker, Jérôme Pety, Johannes Puschnig, Miguel Querejeta, Toshiki Saito, Karin Sandstrom, Eva Schinnerer, Andreas Schruba

%\collaboration{1}{(AAA collaboration)}

% \author{CCC CCC}
% \altaffiliation{CCC, CCC}
% \affiliation{CCC, CCC, CCC, CCC}
% \affiliation{CCC, CCC, CCC, CCC, CCC}
% \nocollaboration{1}

% \author{DDD DDD}
% \affiliation{DDD, DDD, DDD, DDD}
% \nocollaboration{2}

\suppressAffiliations

\begin{abstract}
We compare embedded young massive star clusters (YMCs) to (sub-)millimeter line observations tracing the excitation and dissociation of molecular gas in the starburst ring of NGC~1365. This galaxy hosts one of the strongest nuclear starbursts and richest populations of YMCs within 20~Mpc. 
Here we combine near-/mid-IR PHANGS-JWST imaging with new ALMA multi-$J$ CO(1--0, 2--1 and 4--3) and [\CI](1--0) mapping, which we use to trace CO excitation via $\Remark{R42}=\Remark{ICO43}/\Remark{ICO21}$ and $\Remark{R21}=\Remark{ICO21}/\Remark{ICO10}$ and dissociation via $\Remark{RCICO}=\Remark{ICI10}/\Remark{ICO21}$ at 330~pc resolution. 
We find that the gas flowing into the starburst ring from northeast to southwest appears strongly affected by stellar feedback, showing decreased excitation (lower $\Remark{R42}$) and increased signatures of dissociation (higher $\Remark{RCICO}$) in the downstream regions.
There, radiative transfer modelling suggests that the molecular gas density decreases and temperature and $\XCICO$ abundance ratio increase.
We compare $\Remark{R42}$ and $\Remark{RCICO}$ with local conditions across the regions and find that both correlate with near-IR 2~$\um$ emission tracing the YMCs and with both PAH (11.3~$\um$) and dust continuum (21~$\um$) emission. 
In general, $\Remark{RCICO}$ exhibits $\sim 0.1$~dex tighter correlations than $\Remark{R42}$, suggesting \CI{} to be a more sensitive tracer of changing physical conditions in the NGC~1365 starburst than CO(4--3). 
Our results are consistent with a scenario where gas flows into the two arm regions along the bar, becomes condensed/shocked, forms YMCs, and then these YMCs heat and dissociate the gas. 
\end{abstract}


\section{Introduction}

The interplay between star formation and the interstellar medium (ISM) is a key topic in our understanding of galaxy evolution. Star formation happens in cold molecular gas
\citep[e.g., review by][]{Kennicutt2012Review} and then young massive stars (often found in young massive clusters, or YMCs) drive energetic winds and radiation that may heat up or destroy giant molecular clouds (GMCs), a process known as stellar feedback (e.g., see reviews by \citealt{Chevance2020,Chevance2022}).
An active galactic nucleus (AGN), if present, can also have a significant impact on the host galaxy's ISM and star formation, known as the AGN feedback (see e.g.\ the reviews by \citealt{Cicone2018,Harrison2018}). 
Across the universe, much star formation occurs in gas-rich, turbulent, and high surface density regions \citep[e.g.,][]{Tacconi2020}. To understand how feedback processes operate in such intense environments, the local starburst galaxy population, and especially starbursting galaxy centers, represent key targets where we can observe the physical state of the molecular gas and the impact of stellar feedback in the greatest detail.

Thanks to high resolution optical (e.g., LEGUS: \citealt{LEGUS2015}; PHANGS-HST: \citealt{Lee2022PhangsHst}; PHANGS-MUSE: \citealt{Emsellem2022}; MAD: \citealt{ErrozFerrer2019}; TIMER: \citealt{Gadotti2019}) and mm-wave imaging (e.g., NUGA: \citealt{Combes2019}; PHANGS-ALMA: \citealt{Leroy2021PhangsAlmaSurvey}; GATOS: \citealt{GarciaBurillo2021}), we have made great strides in understanding the interplay of molecular gas, star formation, and stellar feedback in normal galaxy disks (e.g., feedback timescales: \citealt{Grasha2018, Kruijssen2019, Chevance2020, Kim2022, Pan2022}, pressure and turbulence: \citealt{SunJiayi2020, Barnes2021, Barnes2022}; outflows of AGNs: \citealt{Audibert2019,GarciaBurillo2021,Saito2022a,Saito2022b}; to name a few). 
However, the nuclear starbursts, which contain a significant to dominant fraction of host galaxy's star formation and especially the most massive ($M_{\star} \gtrsim 10^{6} \; \mathrm{M_{\odot}}$) YMCs, are still far from being well understood.

NGC~1365 among the most actively star-forming local galaxies ($\mathrm{SFR} = 16.9 \; \mathrm{M_{\odot} \, yr^{-1}}$; \citealt{Leroy2021PhangsAlmaSurvey}) and hosts the richest populations of YMCs in the local $\sim 20$~Mpc Universe \citep{Kristen1997, Galliano2008, WHITMORE_PHANGSJWST}. 
At a distance of $19.57 \pm 0.78$~Mpc (\citealt{Anand2021a,Anand2021b}; $1'' = 95$~pc), its nuclear starbursting ring ($R \sim 1.8$~kpc; \citealt{SCHINNERER_PHANGSJWST}) is fed by gas flowing inward along bar lanes of its well-known 17~kpc-long stellar bar \citep{Lindblad1999}. 
An optical/radio/X-ray AGN is also well-known at its center (e.g., \citealt{Veron1980,Turner1993,Morganti1999,Fazeli2019}).
Based on the first data of PHANGS-JWST (\citealt{LEE_PHANGSJWST}), 37 $M_{\star} \gtrsim 10^{6} \; \mathrm{M_{\odot}}$ and $\mathrm{age} \lesssim 10 \; \mathrm{Myr}$ YMCs are found within the central $\sim 3 \, \mathrm{kpc} \times 2.5 \, \mathrm{kpc}$ area \citep{WHITMORE_PHANGSJWST}, more numerous than in any other galaxy within $\sim 20$~Mpc.

In this work, we utilize the first PHANGS-JWST mid-IR imaging along with new and archival ALMA multi-$J$ (1--0, 2--1 and 4--3) CO and [\CI](1--0) line mapping to assess how tracers of CO excitation, dissociation, and other molecular gas properties relate to the location and likely evolution of embedded YMCs in this rich inner region of NGC~1365.

The spectral line energy distribution (SLED) of CO is a powerful tool to constrain gas temperature and density (e.g., \citealt{Goldreich1974, Israel1995, Bayet2004, Bayet2006, Papadopoulos2007, Papadopoulos2010b, Greve2014, ZhangZhiyu2014, Liudz2015, Rosenberg2015, Israel2015, Kamenetzky2014, Kamenetzky2016, Kamenetzky2017}).
The CO(4--3) transition has more than an order of magnitude higher critical density and upper level energy ($E_{u=4}/k_{\mathrm{B}} = 55.32 \ \mathrm{K}$, $n_{\mathrm{crit.}} \sim 4 \times 10^4 \ \mathrm{cm^{-3}}$; \citealt{Meijerink2007}) than the ground 1--0 transition ($E_{u=1}/k_{\mathrm{B}} = 5.53 \ \mathrm{K}$, $n_{\mathrm{crit.}} \sim 3 \times 10^3 \ \mathrm{cm^{-3}}$)\,\footnote{Although practically these lines can be moderately excited and detectable even at lower densities (e.g., \citealt{Scoville2013Book,Shirley2015}).}. 
Therefore, a higher $\Remark{R41} \equiv \Remark{ICO43}/\Remark{ICO10}$ line ratio generally means a higher density and temperature of gas, but the actual shape of CO SLED (e.g., traced by both a mid-$J$ ratio $\Remark{R42} \equiv \Remark{ICO43}/\Remark{ICO21}$ and a low-$J$ ratio $\Remark{R21} \equiv \Remark{ICO21}/\Remark{ICO10}$) is important to distinguish the effects of changing density, temperature and other ISM properties (e.g., turbulent line width) that may relate to the stellar feedback.

Meanwhile, the \CI{} offers an additional potential diagnostic on the feedback affecting the dissociation of CO. 
\CI{} originates from a thin dissociation layer in photon-dominated regions (PDRs; \citealt{Langer1976,deJong1980,Tielens1985a,Tielens1985b,Hollenbach1991}) exposed to the UV radiation of H~{\sc ii} regions (\citealt{Hollenbach1997,Kaufman1999,Wolfire2022}). 
Cosmic rays and X-rays from YMCs and AGN, known as the cosmic-ray dominated regions (CRDR; \citealt{Papadopoulos2010c,Papadopoulos2011,Papadopoulos2018}) and X-ray dominated regions (XDR; \citealt{Maloney1996,Meijerink2005,Meijerink2007,Wolfire2022}),
respectively, can drastically increase \CI{} abundances and lead to a high \CI{}/CO line ratio ($\Remark{RCICO}$; \citealt{Israel2015,Salak2019,Izumi2020}).

Because embedded YMCs and AGN exert intense radiative and mechanical feedback, they should in principle have a strong effect on the surrounding gas, which should manifest as observable variations in the CO SLED and the \CI{}/CO ratios. 
However, isolating the effects of this feedback is challenging due to the lack of high-resolution, multi-$J$ CO and \CI{} observations and high-resolution, sensitivity mid-IR imaging in the past. 
Therefore, in this Letter, we use the new JWST+ALMA data and present the first multi-$J$ CO and \CI{} + embedded YMC study to trace the molecular gas properties and feedback in the bar-fed central starbursting environment in NGC~1365.

\section{Observations \& Data}

\subsection{JWST observations}

Full descriptions of the PHANGS-JWST observations (Program ID: 02107; PI: J. Lee) are given by \citet{LEE_PHANGSJWST}. 
We provide a brief summary of the observations and processing used for NGC~1365 here. The galaxy was observed during two NIRCam and four MIRI visits. The resulting NIRCam mosaic covers the central $\sim 4.6' \times 2.9'$ ($26.2 \times16.5$ kpc) and the MIRI mosaic covers $\sim 3.8' \times 2.6'$ ($21.6\times14.8$ kpc). 
The NIRCam F200W, F300M and F360M filters cover primarily the stellar continuum. The MIRI F1000W and F2100W filters primarily cover dust continuum imaging. In addition, the NIRCam F335M and MIRI F770W and F1130W filters recover mostly emission associated with Polycyclic Aromatic Hydrogen (PAH) bands in addition to an underlying stellar and dust continuum. Hereafter we refer to the bands by their corresponding wavelengths: 2~$\um$ for F200W, 11.3~$\um$ for F1130W, and 21~$\um$ for F2100W.

We use PHANGS-JWST internal release versions ``v0p4p2'' for NIRCam and ``v0p5'' for MIRI. \citet{LEE_PHANGSJWST} describe the data reduction used for these first results, including modifications to the default JWST pipeline\,\footnote{\url{https://jwst-pipeline.readthedocs.io}}
parameters, 
our customized $1/f$ noise reduction (destriping) and background subtraction for the NIRCam images, the subtraction of MIRI ``off'' images, and how we set the absolute background level in the MIRI images.

JWST's high sensitivity results in saturation of pixels on top of the AGN and the brightest star-forming complexes.
In this work, because we convolve the JWST images to the $3.5''$ resolution of CO(4--3) and [\CI](1--0) lines, the saturation of brightest emission needs to be fixed. 
Here we use a point spread function (PSF) curve-of-growth analysis to match the unsaturated outskirts of each bright saturated spot, i.e., the center of NGC~1365 (where the AGN is located) and the three bright star-forming complexes to the north.

Next, these images are PSF-matched to our common resolution of $3.5''$ corresponding to the ALMA CO(4--3) beam in this work. 
We use the \incode{WebbPSF} software to generate the PSFs and the \incode{photutils} software to create convolution kernels (with fine-tuned TopHatWindow). The images are then resampled with the \incode{reprojection} software in Python to the same world coordinate system and pixel scale.

\begin{figure*}[htb]
    \centering
    \includegraphics[width=\textwidth]{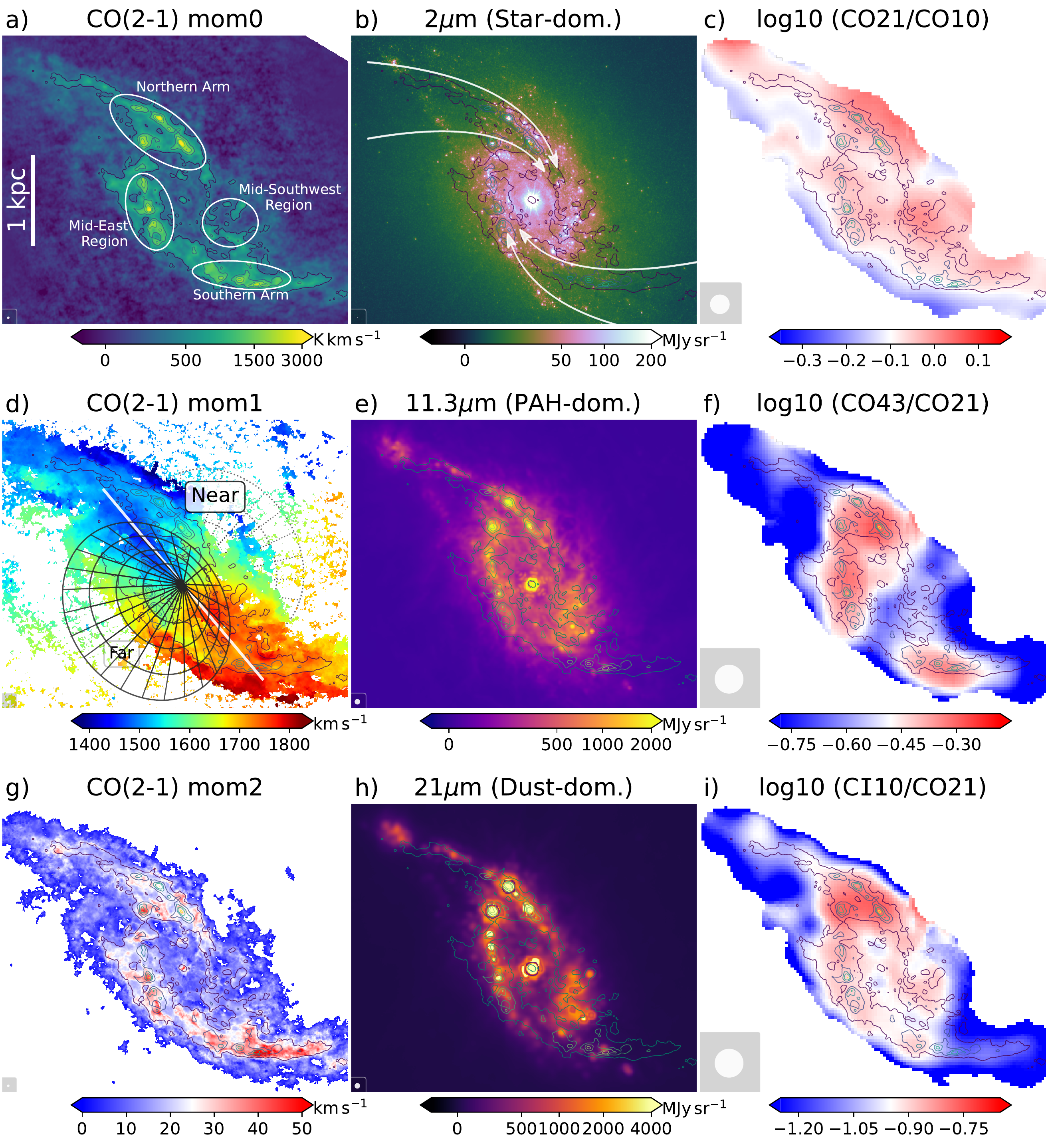}
    \caption{ALMA and JWST images of NGC~1365. 
    \textit{Left column (a, d, g):} Native-resolution ($0.3''$) ALMA CO(2-1) integrated intensity (mom0), line-of-sight velocity (mom1) and velocity dispersion (mom2); see also \citet{SCHINNERER_PHANGSJWST}.
    \textit{Middle column (b, e, h):} JWST stellar continuum dominated NIRCam 2~\um, PAH-dominated MIRI 11.3~\um, and dust continuum dominated MIRI 21~\um, all at native resolutions; see also \citet{WHITMORE_PHANGSJWST}. 
    \textit{Right column (c, f, i):} Ratio maps in logarithmic scale for $\Remark{R21} = \Remark{ICO21}/\Remark{ICO10}$, $\Remark{R42} = \Remark{ICO43}/\Remark{ICO21}$ and $\Remark{RCICO} = \Remark{ICI10}/\Remark{ICO21}$ at the common resolution of $3.5''$ (beam is shown in bottom left). 
    Same contours of $\Remark{ICO21}$ at 400, 1000, 1500, 2000 and 3000 $\Remark{Kkms}$ are shown in all panels. 
    Four regions mentioned in this work are highlighted in panel \textit{a)}. 
    The cone in panel \textit{d)} represents the known 
    ionized gas outflow with a half-opening angle $50^\circ$, cone axis inclination $35^{\circ}$ and PA $126^{\circ}$ (\citealt{Hjelm1996}; see also \citealt{Sandqvist1995,Lindblad1999,Venturi2018,GaoYulong2021}) and the white line indicates the disk kinematic major axis at PA $220^\circ$ (\citealt{Jorsater1995,Lindblad1999}). 
    }
    \label{fig:1}
\end{figure*}

\subsection{ALMA observations of CO(1--0), (2--1) and (4--3), and [\CI](1--0)}
\label{subsec:alma}

We have obtained new ALMA ACA (7m) Band 8 observations (program 2019.1.01635.S, PI: D. Liu) that map the CO(4--3) and [\CI](1--0) from the inner part of NGC~1365. The CO(4--3) observations at 458.53~GHz cover the inner $75'' \times 22''$ ($7.3 \times 2.1$~kpc$^2$) using a 19-pointing mosaic and on-source integration time of $\sim 1$~hour and had $T_{\mathrm{sys}} \sim 550$~K. The [\CI](1--0) observations at 489.48~GHz covered the same area using a 23-pointing mosaic and an on-source integration time of $\sim 8$~hours with $T_{\mathrm{sys}} \sim 500 \text{--} 1000$~K. 
The raw data were reduced using the standard ALMA calibration pipeline with CASA software \citep{CASA}.

We also reduced archival ALMA 12m+7m observations of CO(1--0) (2015.1.01135.S, PI: F. Egusa; 2017.1.00129.S, PI: K. Morokuma), 12m+7m observations of CO(2--1) (2013.1.01161.S, PI: K. Sakamoto), 
including the CO(1--0) and (2--1) total power (TP) data.

After running the ALMA pipeline, we imaged and processed our data into data cubes using the PHANGS--ALMA pipeline\,\footnote{\url{https://github.com/akleroy/phangs_imaging_scripts}} \citep{Leroy2021PhangsAlmaPipeline}. 
Then we used the \incode{spectral_cube}\footnote{\url{spectral-cube.readthedocs.io}} software to convolve our cubes to the common beam of $3.5''$ matching the coarsest \CI{} data. 

The native resolution of the CO(1--0) and (2--1) data are $2.4''$ and $0.31''$. Based on the arrays used, these images have maximum recoverable scales (MRS) of $\sim 80''$ and $40''$, respectively.
Our CO(4--3) and [\CI](1--0) data have $3.4 \text{--} 3.5''$ beams and $15 \text{--} 20''$ MRSs. 
For CO(1--0) and (2--1) the short-spacing is corrected with the PHANGS--ALMA pipeline using their total power data, however, other lines lack the short-spacing correction. 
We considered using the archival 12m-only ALMA CO(3--2) data for this galaxy but found that its poor $u-v$ coverage made it not useful in the analysis presented.

We expect only a minor percentage of missing flux for our CO(4--3) and [\CI](1--0) data (e.g., $\lesssim 25\%$). 
Indeed, our total CO(4--3) line luminosity within a radius $\sim 14''$ is $5.3 \times 10^{8} \ \mathrm{K \, km \, s^{-1}\, pc^{2}}$, which is more than half of the total \textit{Herschel} SPIRE FTS CO(4--3) luminosity within a $\sim 43''$ beam ($7.3 \times 10^{9} \pm 1.2 \times 10^{7} \ \mathrm{K \, km \, s^{-1}\, pc^{2}}$; \citealt{Liudz2015}; scaled to the same distance). 
Simulations of visibilities mimicking similar ACA Band 8 observations also reveal that the missing flux is minimal at relatively bright line emission spots ($\lesssim 10\%$; D. Liu et al. submitted).

We collapsed the data cubes into moment maps using the PHANGS-ALMA imaging and post-processing pipeline. We specifically focus on the ``strict,'' high-confidence mask, which is constructed using a watershed algorithm with relatively high clipping values (\citealt{Rosolowsky2006,Rosolowsky2021}) and so expected to include only significant detections. 
We build a single common strict mask by combining the individual masks for CO(1--0), CO(2--1), CO(4--3) and $\Remark{CI10}$, then we extracted moment maps for each line therein. The error maps are also measured from the data cube, then computed for moment maps following the standard procedures implemented in the PHANGS--ALMA pipeline. 
In our final moment-0 maps, the S/N is 5--200 for CO(4--3), 5--100 for $\Remark{CI10}$, and much higher (20--250) for CO(1--0) and CO(2--1). 
We create line ratio maps from every pair or lines at the best common resolution of $3.5''$ and use them to compute correlations and CO SLEDs.

\subsection{Auxiliary data products and complementary information}
\label{subsec:auxdata}

We use the YMC catalog from \citet[][this Issue]{WHITMORE_PHANGSJWST} to determine the locations where CO and \CI{} line fluxes are extracted and studied. 
It contains 37 YMCs with HST or JWST photometry derived masses $M_{\star} \sim 1 \times 10^{6} \ \text{--} \ 2 \times 10^{7} \ \mathrm{M_{\odot}}$ and ages $\sim 1 \ \text{--} \ 8 \ \mathrm{Myr}$.

We use the CO(2--1) line-of-sight decomposition and orbital time information from \citet[][this Issue]{SCHINNERER_PHANGSJWST} for discussions relating to line velocity dispersion. 
\cite{SCHINNERER_PHANGSJWST} present the $0.3''$-resolution CO(2--1) data and decomposition of each line-of-sight spectrum into multiple Gaussian components with the \incode{ScousePy} software \citep{Henshaw2016,Henshaw2019}. 
They find that multiple velocity components exists in the system, and individual components have a typical line width $\sigma \sim 19 \ \mathrm{km \, s^{-1}}$. 
We use this value as the microturbulent line width $\sigma$ for our later radiative transfer calculation. 
They derive a dynamical timescale $t_{\mathrm{dyn}} \sim 20 \ \mathrm{Myr}$ at $R = 5'' \approx 475 \, \mathrm{pc}$, which is roughly the radius of the dynamically cold inner gas disk in this starburst ring system. 
This time scale is used as a context for discussion in this work.

\section{Results \& Discussion}

\subsection{Spatial and kinematic structure}
\label{subsec:3.1}

In Fig.~\ref{fig:1} we present the ALMA and JWST data at their native resolution ($\sim 6$~pc at 2~\um, $\sim 30$~pc for CO(2--1) and $\sim 60$~pc at 21~\um) and the multi-$J$ CO and \CI{} ratio maps at their common resolution of $3.5''$ ($\sim 330$~pc). 
We define $\Remark{R21} = \Remark{ICO21}/\Remark{ICO10}$ to trace the low-$J$ CO excitation, $\Remark{R42} = \Remark{ICO43}/\Remark{ICO21}$ for the mid-$J$ excitation, and $\Remark{RCICO} = \Remark{ICI10}/\Remark{ICO21}$ as a tentative tracer of CO dissociation.

We also label four regions in this study to facilitate the later discussion: 
\textit{i)} the ``Northern Arm'', which is where the gas flows in from northeast along the bar; 
\textit{ii)} the ``Mid-Southwest'' region, which is the downstream location of the gas coming from the Northern Arm (and we call the Northern Arm the upstream of the Mid-Southwest region); 
\textit{iii)} the ``Southern Arm'', similar to the Northern Arm, is the location of gas flowing in from the southeast along the bar; 
\textit{iv)} the ``Mid-East'' region, considered as the downstream of the Southern Arm. 
The Northern Arm and Mid-Southwest both belong to the ``northern bar lane'' as defined in \cite{SCHINNERER_PHANGSJWST}, and the Southern Arm and Mid-East correspond to the ``southern bar lane'' therein. 
The Northern Arm alone is also emphasized as the ``region 1'' in \cite{WHITMORE_PHANGSJWST}.

\begin{figure}[tb]
\centering
\includegraphics[width=\linewidth]{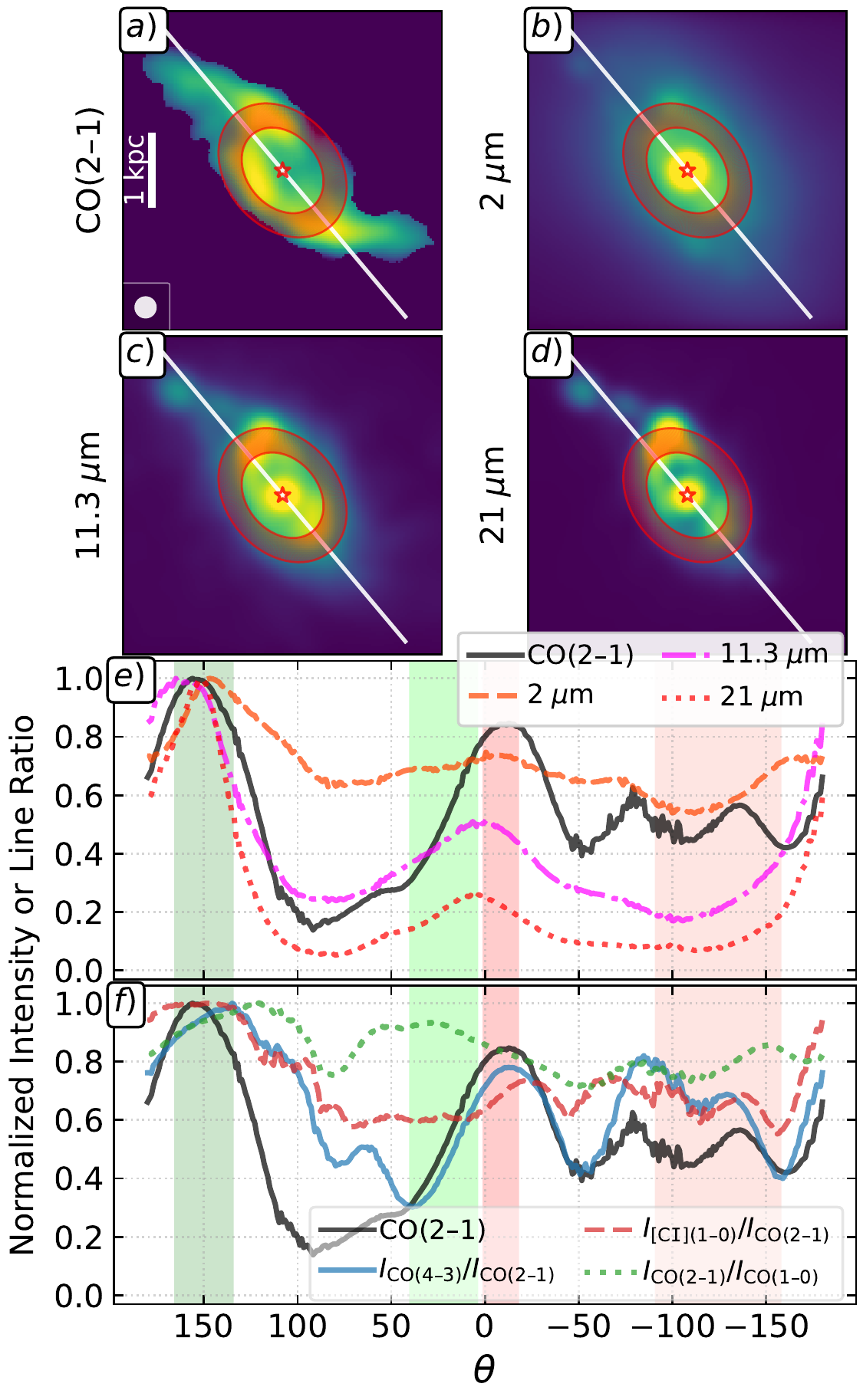}
\caption{%
\textit{Panels a)--d)} are the CO(2--1) line integrated intensity, 2~\um{}, 11.3~\um{} and 21~\um{} images, PSF-matched to our $3.5''$ resolution. A common annulus is shown in all panels with radius~$=0.9 \pm 0.2$~kpc, inclination $40^{\circ}$ and PA $220^{\circ}$ \citep[white solid line][]{Lindblad1999}. 
\textit{Panels e)--f)} show the normalized azimuthal profiles of the four upper panels and $\Remark{R42}$, $\Remark{RCICO}$ and $\Remark{R21}$ line ratios as a function of azimuthal angle $\theta$ measured from the PA of the receding side of kinematic major axis. 
Dark green, light green, dark red and light red shading represents the Northern Arm, Mid-Southwest, Southern Arm and the Mid-East regions, respectively. 
\label{fig:asymmetry}
}
\end{figure}

The CO emission mostly arises from the prominent starburst ring, oriented from northeast to southwest, but exhibits a highly asymmetric distribution within this ring. 
The 2\,\um\, stellar emission, which is much less affected by dust attenuation than HST optical images, reveals a more symmetric distribution than CO. 
We show the azimuthal profiles of the CO, 2~\um, 11.3~\um\ and 21~\um\ emission along the starburst ring in Fig.~\ref{fig:asymmetry}. 
These profiles are measured from the radially-averaged emission at each azimuthal angle $\theta$ using a common annulus at our working resolution of $3.5''$. 
Here, $\theta = 0^{\circ}$ corresponds to the receding side of major kinematic axis with PA of $220^{\circ}$, and $\theta = 180^{\circ}$ ($-180^{\circ}$) to the approaching side. 
All azimuthal profiles show a peak at $\theta_{\mathrm{max}} \sim 150^{\circ}$, corresponding to the Northern Arm region. 
The 2~\um\ profile shows the least azimuthal variation of about 50\% (with a maxima-to-minima ratio of 1.9), whereas the other three profiles show large variations $\sim 80\% \text{--} 95\%$, with a maxima-to-minima ratio of 8.5, 6.0 and 19.4, for CO, 11.3~\um\ and 21~\um\, respectively.

The Northern Arm peak ($\theta \sim 150^{\circ}$) in the azimuthal profiles mainly consists of three extremely massive star clusters, YMC~29, 33 and 28, corresponding to \cite{Galliano2005,Galliano2008} mid-IR ID M4, M5 and M6, and \cite{Sandqvist1995} radio 20~cm/6~cm ID D, E and G, respectively.

\begin{figure}[tb]
\centering
\includegraphics[width=\linewidth]{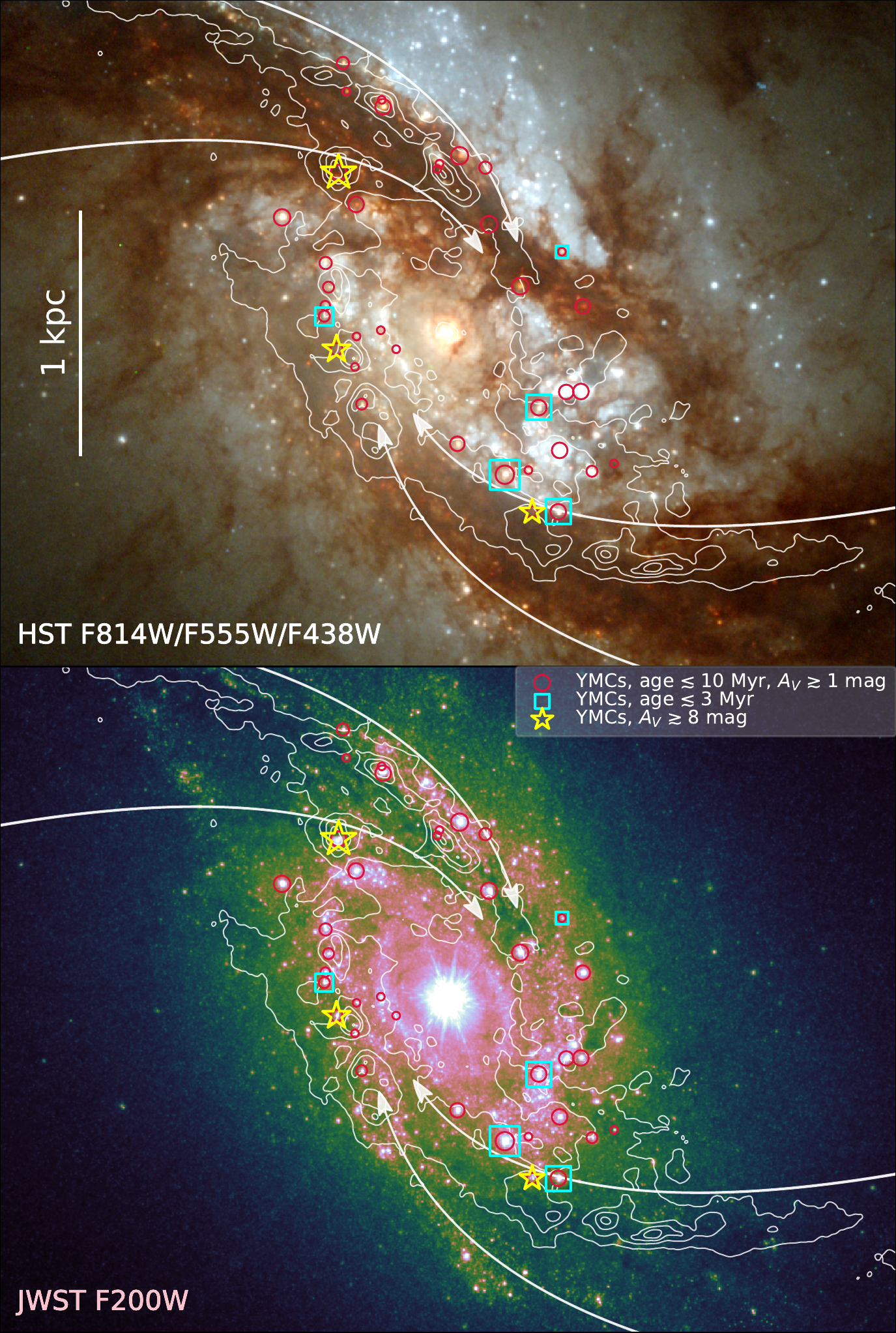}\\[2ex]
\caption{%
YMCs \citep{WHITMORE_PHANGSJWST} overlaid on HST RGB (F814W/F555W/F438W) image (\textit{top}) and JWST F200W (2\,\um) image (\textit{bottom}). 
All YMCs are shown as red circles. 
Symbol sizes from smallest to largest indicate masses from $\sim 10^{6.0} \; \Msun$ to $\sim 10^{7.1} \; \Msun$. 
Cyan squares are YMCs with an age less than $\sim 3$~Myrs, and yellow stars are YMCs with an attenuation $A_V > 8$~magnitude. 
We show these YMCs for the context of the discussion in Sect.~\ref{subsec:3.1} of this work. 
See \citet{WHITMORE_PHANGSJWST} for the original and detailed study of these YMCs. 
\label{fig:YMCs}
}
\end{figure}

In Fig.~\ref{fig:YMCs} we show the distribution of YMCs on the HST RGB and JWST 2~\um\ image and indicate the dust lanes by two pairs of curved arrows. 
The CO(2--1) emission contours broadly match the dust lanes in the Northern and Southern Arm regions but appear to be narrow at the edge of dust lanes. 
Such dust lanes in barred galaxies are usually seen as a result of dissipative processes in the gaseous component, and the widely accepted view is that they delineate shocks in the interstellar gas (e.g., see reviews by \citealt{Sellwood1993,Buta1996,Lindblad1999}; see also \citealt{Pastras2022} for detailed simulations). 

In addition to the gas and dust lanes, 
we also highlight in Fig.~\ref{fig:YMCs} the YMCs younger than $\sim 3$~Myr and more attenuated than $A_V \sim 8$~magnitude with cyan squares and yellow stars, respectively.
The majority of the most massive, young and attenuated YMCs sit in the Mid-Southwest and Mid-East regions.

The multi-$J$ CO and \CI{} line ratio maps exhibit a significant asymmetry as well. The Northern Arm region shows a factor of 1.5--2$\times$ higher excitation and dissociation than other regions, as traced by \Remark{R42} and \Remark{RCICO}. 
In particular, the \Remark{R42} and \Remark{RCICO} peaks coincide with the YMC ID~29 in the Northern Arm, which also corresponds to the mid-IR source M4 in \cite{Galliano2005,Galliano2008}. 
It has the highest CO excitation ($\Remark{R42} \sim 0.51 \pm 0.01$, $\Remark{R21} \sim 0.92 \pm 0.01$) and shows the most signs of dissociation ($\Remark{RCICO} \sim 0.17$). 
The distributions of low-$J$ excitation, \Remark{R21}, mid-$J$ excitation \Remark{R42}, and our photodissociation tracer \Remark{RCICO} also appear distinct from one another.
\Remark{R21} shows a more symmetric, smooth distribution than the other two line ratios. The $\log \Remark{R21}$ is above $-0.1$ in most regions, indicating that CO(2--1) is almost thermalized in the regions we study. The \Remark{R42} and \Remark{RCICO} ratios have much larger dynamical ranges, and both peak in the Northern Arm. However, \Remark{R42} is depressed whereas \Remark{RCICO} is enhanced in the Mid-Southwest region, and their trends are the opposite in the Southern Arm region. 

The molecular gas has complex, rapidly-changing kinematics in these regions. Panels (d) and (g) in Fig.~\ref{fig:1} show the line-of-sight velocity and line velocity dispersion maps. 
The Northern and Southern Arm regions correspond to locations where molecular gas flows inward along the bar. The dissipative gas flow in such a strong bar potential can trigger shocks at the leading edge of the bar, appearing as dust lanes following the orbital skeleton (e.g., \citealt{Athanassoula1992,Sellwood1993,Buta1996,Lindblad1999,Sellwood2022,Pastras2022}; see also Fig.~1 of \citealt{Maciejewski2002}). 
In the Southern Arm, the enhanced \Remark{R42}, non-enhanced \Remark{RCICO}, 
lack of YMCs and elevated velocity dispersion $\sigma \sim 50 \ \mathrm{km \, s^{-1}}$ (Fig.~\ref{fig:1}\,g)
may suggest a scenario where shocks triggered at the edge of dust lanes are compressing the gas and hence highly-exciting the mid-$J$ CO, but not leading to significant CO dissociation or higher \CI{} abundance in this region. 
However, in other cases, e.g., the Northern Arm, when the gas is so dense and even fragments to trigger intensive star formation and hence stellar feedback, the effect of shocks 
can be mixed with the stellar feedback effect.
Future observations of shock tracers like SiO or HNCO for less destructive shocks \citep[e.g.][]{Usero2006,GarciaBurillo2010,Meier2015,Kelly2017} will be the key to shed light on this shock scenario.

In the panel~\textit{(d)} of Fig.~\ref{fig:1} we also highlight the well-known ionized gas outflow \citep{Phillips1983,Hjelm1996,Lindblad1999,Sandqvist1995,Sakamoto2007,Lena2016,Venturi2018,GaoYulong2021}. 
We do not find a high $\Remark{R42}$ nor enhanced $\Remark{RCICO}$ peaking at the center and extending along the outflow cone, unlike the cases of known \CI\ molecular outflows in AGN host galaxies (e.g., \citealt{Saito2022a,Saito2022b}; with $\Remark{RCICO} \sim 1$).
This agrees with \citet{SCHINNERER_PHANGSJWST} who find no evidence for the ionized outflow to intersect the molecular gas disk, and with \cite{Combes2019} who find that this region lacks a molecular gas outflow (and instead shows evidence for inflow).

\begin{figure*}[htb]
\centering
\includegraphics[width=\textwidth]{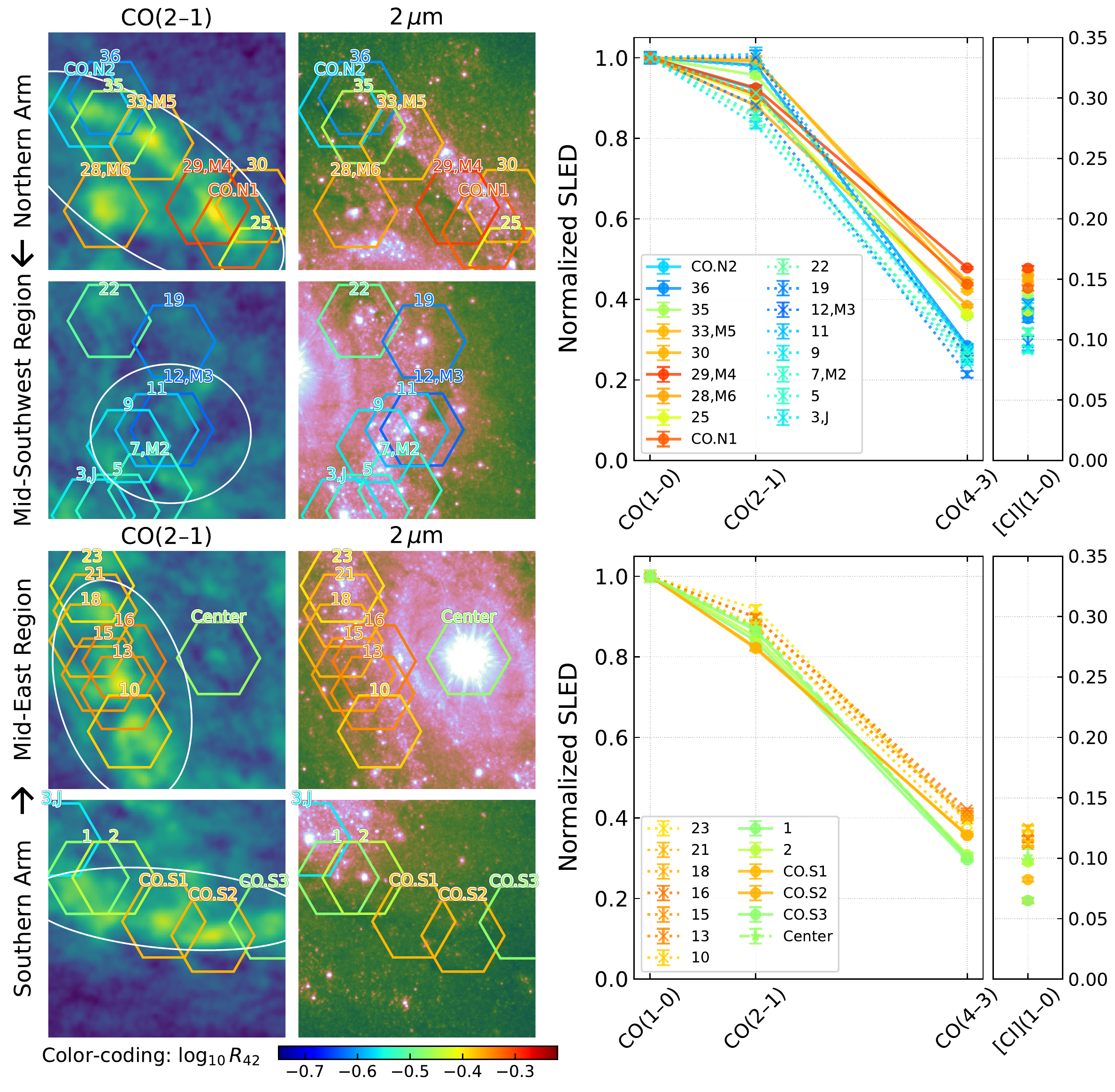}
\caption{%
    \textit{Left panels:} CO(2--1) and 2~$\um$ zoom-in images of the Northern Arm, Mid-Southwest, Mid-East and Southern Arm regions (from \textit{top} to \textit{bottom}). 
    Hexagons indicate YMCs identified by \cite{WHITMORE_PHANGSJWST} (with numbers indicating their ID from that paper), YMC-free CO peaks (labelled ``CO.N1--N2'' and``CO.S1--S3'' in Northern and Southern regions, respectively, visually identified from this work), and the galaxy center. 
    The ``M2''--``M6'' labels indicate YMCs that have a mid-IR counterpart from \cite{Galliano2005}. 
    The ``J'' label indicates that YMC~3 has a radio counterpart from \cite{Sandqvist1995}. 
    All images have the same field of view of $10''$ ($\sim 950$~pc) and color scaling as in Fig.~\ref{fig:1}. 
    Hexagons have the same diameter of $3.5''$ ($\sim 330$~pc). 
    The black arrows at left indicate the up-to-down-stream directions. 
    \textit{Right panels}: CO and \CI{} SLEDs of the hexagons centered at YMCs or YMC-free CO peaks in the corresponding upper/lower four left image panels. 
    Hexagons and SLEDs are color-coded by their \Remark{R42} (see the bottom color bar).
    Solid and dotted lines represent up- and downstream regions, respectively. 
    The Center's CO SLED is shown as the dashed line in the lower panel. 
    These CO and \CI\ SLEDs are used for the radiative transfer modeling fitting in Sect.~\ref{subsec:zoom-in} with gas properties reported in Table~\ref{tab:2}.
}
\label{fig:zoom-in}
\end{figure*}

\subsection{Zoom-in views and CO and \CI\ excitation}
\label{subsec:zoom-in}

We examine the aforementioned regions in more detail in Fig.~\ref{fig:zoom-in}. 
We first draw hexagonal apertures centered on bright YMCs \citep{WHITMORE_PHANGSJWST}, 
CO peaks that do not coincide with a YMC (based on Fig.~\ref{fig:zoom-in}, hereafter YMC-free CO peaks, labeled as CO.N1/N2 and CO.S1/S2/S3 for those in the Northern and Southern Arms, respectively), and the galaxy center, each with a 330~pc diameter. 
Then, we take the 
pixel value at the hexagon center from the $3.5''$-resolution moment-0 map for each line, and show their normalized CO and \CI{} SLED in the 
right panels of Fig.~\ref{fig:zoom-in}. 
The hexagons in the 
left panels are then color-coded by their mid-$J$ excitation $\Remark{R42}$. 
In this way we can clearly see how gas in the apertures is excited. 

The most (mid-$J$) excited aperture is YMC~29 (M4) coinciding with the brightest CO peak in the Northern Arm, with $\Remark{R42} = 0.50 \pm 0.01$ (and $\Remark{R21} = 0.92 \pm 0.01$, $\Remark{R41} = 0.46 \pm 0.01$).
Its stellar age is estimated as $\sim 3$~Myr (with a $\sim 0.4$--0.6~dex uncertainty; \citealt{WHITMORE_PHANGSJWST}). 
The least (mid-$J$) excited aperture we analyzed is YMC~12 (M3) in the Mid-Southwest region, with $\Remark{R42} = 0.24 \pm 0.02$. However, its low-$J$ excitation is still quite high, with 
$\Remark{R21} = \ 0.98 \pm 0.02$. 
It is also very young, with a stellar age $\sim 3$~Myr (with a similarly large uncertainty; \citealt{WHITMORE_PHANGSJWST}).
Other YMCs and YMC-free CO peaks have $\Remark{R42}$ values in-between these two extreme cases.

Based on the motion of gas along the dust lanes and starburst ring (Figs.~\ref{fig:1}--\ref{fig:YMCs}), the Northern Arm is upstream of the Mid-Southwest region, and the Southern Arm is upstream of the Mid-East region. Gas enters the system first via the Northern/Southern Arms then moves to the Mid-Southwest/Mid-East regions, forming the starburst ring. 
Along the Northern Arm, gas moves from the less-excited CO.N2 location to the highly-excited CO.N1 position. Then, in about a quarter of the orbital time, e.g., $\sim 5$~Myr, the gas will arrive in the Mid-Southwest region, where the CO excitation is decreasing again. It is expected that gas will move from the Mid-Southwest region towards the Southern Arm and possibly trigger collision and compression. 
Similarly, starting from the Southern Arm, the gas moves from CO.S3 to CO.S2 and CO.S1 positions, then circles to the Mid-East region, and heads towards the Norther Arm. 
The youngest YMCs are found in these possibly colliding areas, i.e., YMC~3 (J) and YMC~28 (M6, G), which both have an age $\sim 1$~Myr and are highlighted as yellow stars in Fig.~\ref{fig:YMCs}.\,\footnote{%
This gas colliding scenario is similar to the hypothesis of the formation of the youngest super star cluster RCW~38 in the Milky Way \citep{Fukui2016}.
}

\begin{figure*}[htb]
\centering
\includegraphics[width=\textwidth]{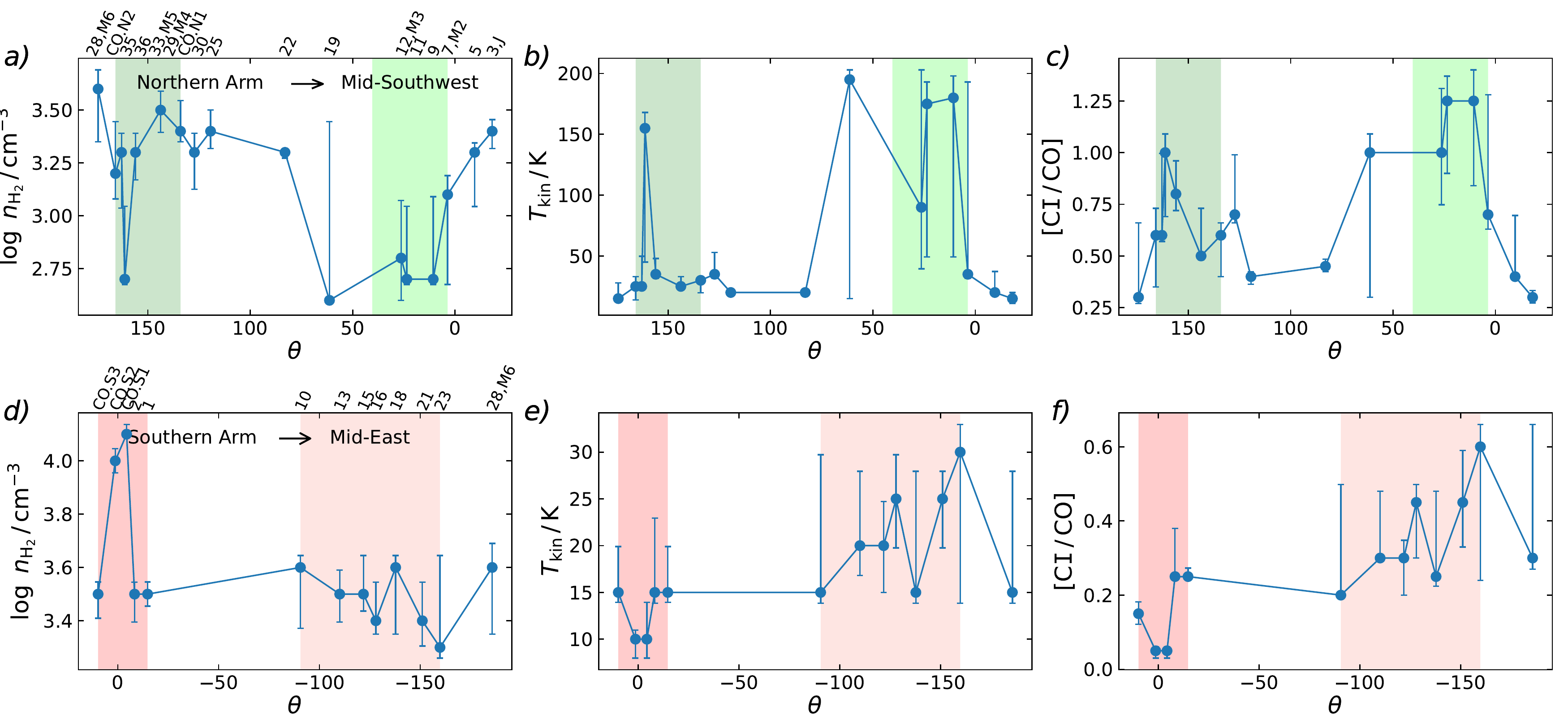}
\caption{%
Non-LTE LVG model fitting results for CO and \CI\ SLEDs in hexagon apertures (see Fig.~\ref{fig:zoom-in}) in the Northern Arm to Mid-Southwest regions (\textit{upper panels}) and Southern Arm to Mid-East regions (\textit{lower panels}), respectively. 
The $x$-axis is the azimuthal angle starting clockwise from the PA~$=220^{\circ}$ direction, same as in Fig.~\ref{fig:asymmetry}. 
Dark/light green and red shadings indicate the $\theta$ range of the corresponding regions. 
The $y$-axes show (\textit{from left to right}) gas volume density ($\lgnHtwo$), kinetic temperature ($\Tkin$) and the $\XCICO$ abundance ratio. 
Hexagon IDs are indicated at the top of the \textit{left} panels at their corresponding $\theta$. 
Error bars are the $\pm 1 \sigma$ uncertainty from the fitting (based on \incode{MICHI2}; \citealt{Liudz2021}; see also Appendix~\ref{appendix:LVG}). 
We discuss the trends of decreasing $\lgnHtwo$ and increasing $\Tkin$ and $\XCICO$ from the upstream (dark green/red, left part of each panel) to the downstream regions (light green/red, right part of each panel) in Sect.~\ref{subsec:zoom-in}. 
\label{fig:LVG}
}
\end{figure*}

To obtain a quantitative description of the gas excitation and photodissociation along the starburst ring,
we perform a non-local thermodynamic (non-LTE) large velocity gradient (LVG) radiative transfer modeling of the 
CO and \CI\ SLEDs for the YMC apertures and CO peaks to evaluate the gas density and temperature and $\XCICO$ abundance ratio. 
We use the \incode{MICHI2} Monte Carlo fitting tool (\citealt{Liudz2021}), with templates generated by the \incode{RADEX} software \citep{vanderTak2007} with a grid of gas kinetic temperature 
$T_{\mathrm{kin}} \sim 5 \textit{--} 200 \ \mathrm{K}$, 
H$_2$ volume density ($\log n_{\mathrm{H_2}} / \mathrm{cm^{-3}} = 2.0 \text{--} 5.0$), 
$\XCICO$ abundance ratio 0.05--3.0, and a fixed line turbulence FWHM 
${\Delta v} = 45 \ \mathrm{km \, s^{-1}}$ (matching the average CO line width of $\sigma \sim 19 \ \mathrm{km \, s^{-1}}$ inferred at 30~pc resolution; \citealt{SCHINNERER_PHANGSJWST}; and Fig.~\ref{fig:1}), as well as a free beam filling factor\,\footnote{A fixed CO abundance per velocity gradient $\mathrm{[CO/H_2]/(\mathrm{d} v/\mathrm{d} r)} = 10^{-5} \ \mathrm{km^{-1} \, s \, pc}$ is also adopted (e.g. \citealt{Curran2001NGC4945,Weiss2007,ZhangZhiyu2014,Liudz2021}). This leads to a non-independent CO column density $N_{\mathrm{CO}} = \mathrm{[CO/H_2]} \times n_{\mathrm{H_2}} \times {\Delta v}/(\mathrm{d} v/\mathrm{d} r)$ that has reasonable values from our fitting ($\log N_{\mathrm{CO}} / \mathrm{cm^{-2}} \sim 17.5 \text{--} 18.7$).}. 
We show examples of our LVG fitting to the most and least excited YMC apertures, YMC~29 (M4) and YMC~12 (M3), respectively, in Appendix~\ref{appendix:LVG}. 
We find $\log (n_{\mathrm{H_2}} / \mathrm{cm^{-3}}) \sim 3.50^{+0.09}_{-0.11}$ and $\sim 2.80^{+0.27}_{-0.20}$, and $T_{\mathrm{kin}} \sim 25^{+8}_{-2}$~K and $\sim 90^{+113}_{-51}$~K, respectively. 
Table~\ref{tab:2} reports the best-fit parameters and 1-$\sigma$ uncertainties for all apertures.

In Fig.~\ref{fig:LVG} we present the LVG fitting results for all apertures whose CO plus \CI\ SLEDs are shown in Fig.~\ref{fig:zoom-in}. 
The fitted $\lgnHtwo$, $\Tkin$ and $\XCICO$ are plotted as functions of azimuthal angle $\theta$ (see Fig.~\ref{fig:asymmetry}). 
In the upper panels of Fig.~\ref{fig:LVG}, $\theta$ ranges from $\sim 160^{\circ}$ to $\sim 10^{\circ}$, in clockwise direction tracking the gas movement from the Northern Arm (dark green shading) to the Mid-Southwest region (light green shading). 
Similarly, in the lower panels of Fig.~\ref{fig:LVG}, $\theta$ spans $\sim 0^{\circ}$ to $\sim -150^{\circ}$, in clockwise direction from the Southern Arm (dark red shading) to the Mid-East region (light red shading). 
We find tentative trends that $\lgnHtwo$ is lower, $\Tkin$ is higher and $\XCICO$ is higher downstream the gas flow. 
The $\XCICO$ ratio increases up to $\sim 1.4$ at the Mid-Southwest YMC~13, YMC~12 (M3) and YMC~10 locations. 
Such a high \CI\ abundance relative to CO is likely due to the strong radiation field created by the YMCs in a relatively diffuse molecular gas. 

Therefore, we propose a scenario where gas arriving from the dust lanes is piled up, compressed and shocked at the Arm regions, and YMCs are formed; 
then gas and YMCs travel to the downstream region during about a quarter of the orbital period (a few Myrs); 
the gas may continue forming stars (clusters) on the way, but must undergo strong stellar feedback, so that it is heated and \CI{}-enriched upon its arrival in the Mid-Southwest/Mid-East regions, as our LVG fitting results suggest.%

We caution that the above picture still needs higher-resolution observational supports. The trends are only significant in the Northern Arm to Mid-Southwest side ($\Tkin$ increases from below $50$~K to above $100$~K and $\XCICO$ increases by a factor of two). The other side of the starburst ring from Southern Arm to Mid-East region indeed has much weaker temperature and $\XCICO$ variations. The variations of $\nHtwo$, $\Tkin$ and $\XCICO$ along the ring are also highly non-monotonous. 
We note that the environment in this starburst ring system is highly dynamic and stochastic. 
The density of the inflow cold gas feeding the starburst ring likely has an impact on how much the stellar feedback can affect the natal molecular gas. 
The Southern Arm gas density is much higher than that of the Northern Arm from our fitting, in line with the gas in the Southern Arm/Mid East regions being more shielded and less affected by the stellar feedback.

Finally, although the galaxy center hosts a Seyfert~1.5 AGN and a prominent [O~\textsc{iii}] ionized gas outflow (\citealt{Phillips1983, Lindblad1999, Sandqvist1995, Hjelm1996, Venturi2018, Sakamoto2007, GaoYulong2021}), it shows only moderately excited CO and \CI{}. As shown in the next section, the center's line ratios tracing CO excitation and dissociation are consistent with other regions when correlating these line ratios with mid-IR emission. At our resolution, we do not find any evidence of an extreme XDR such as a highly-excited CO SLED as seen in, e.g., Mrk~231 (\citealt{vanderWerf2010}), NGC~1068 (\citealt{Spinoglio2012}), and other local ultra-luminous IR galaxies whose global $\Remark{R41}$ are highly-excited or even close to being thermalized ($\sim 0.85 \text{--} 1.1$; see also \citealt{Rangwala2011,Meijerink2013,Kamenetzky2014,Kamenetzky2016,Kamenetzky2017,Glenn2015,Liudz2015,Rosenberg2015,LuNanyao2017}). 
We observe $\Remark{ICI10}/\Remark{ICO10} = 0.10$ and $\Remark{ICI10}/\Remark{ICO21} = 0.11$ in NGC~1365's center at 330~pc-resolution. The fitted $\XCICO$ is $0.25^{+0.23}_{-0.01}$ (Table~\ref{tab:2}). 
These line ratios and abundance ratio are much lower than those measured in the more powerful AGNs in NGC~7469 \citep{Izumi2020} and NGC~1068 \citep{Saito2022a,Saito2022b}, which have $\Remark{ICI10}/\Remark{ICO10} \sim 0.5 \text{--} 1$ and $\XCICO \gtrsim 1$.

\begin{figure*}[htb]
    \centering
    \includegraphics[width=\textwidth]{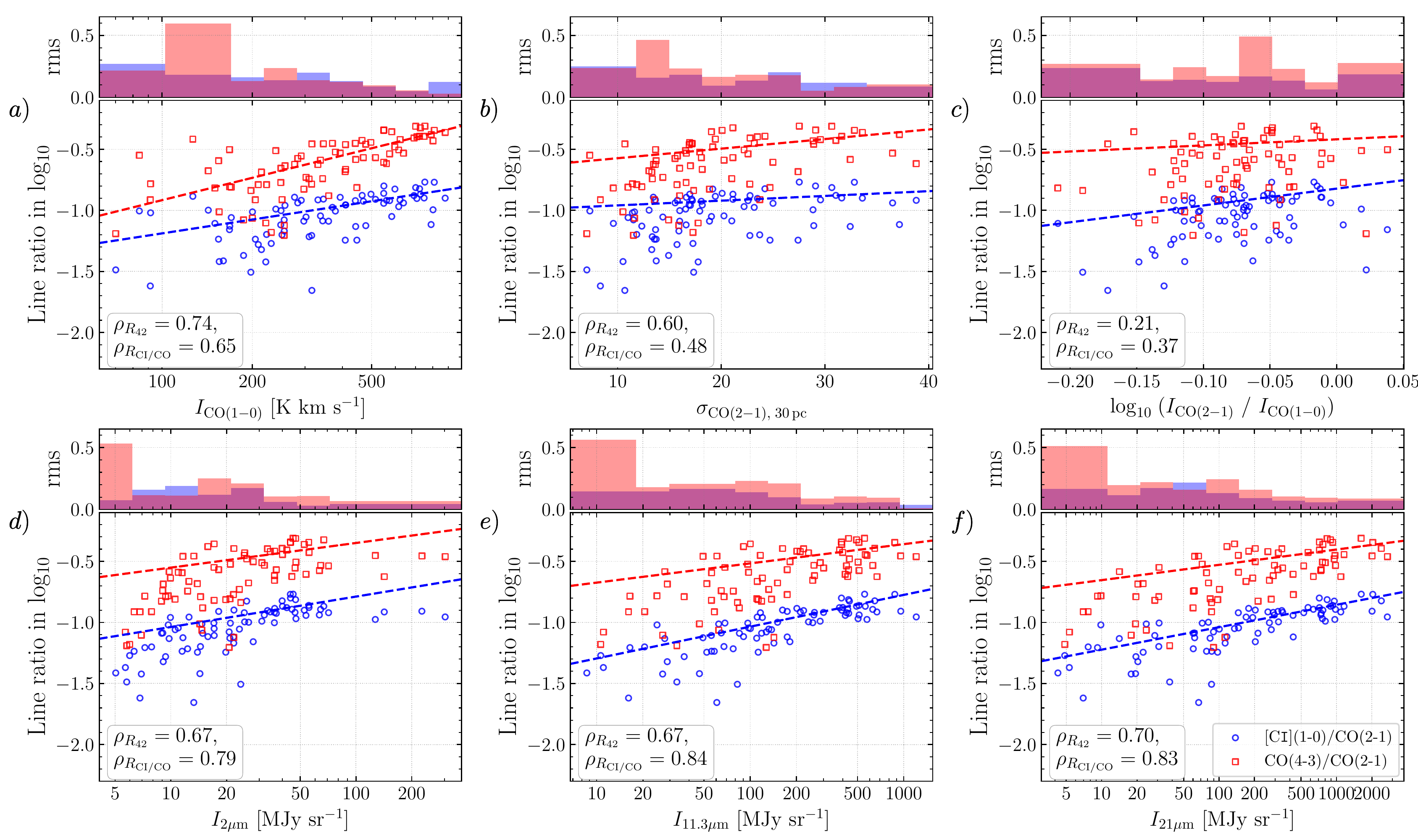}
    \caption{%
    \textit{Panels a) to f):} \Remark{R42} in red and \Remark{RCICO} in blue
    as functions of six variables at 330~pc resolution across the inner part of NGC~1365:
    \textit{a)}  
    CO(1--0) line-integrated intensity, 
    \textit{b)} velocity dispersion, 
    \textit{c)} \Remark{R21} line ratio, 
    \textit{d)} 2~$\um$, 
    \textit{e)} 11.3~$\um$, and 
    \textit{f)} 21~$\um$ emission. 
    Dashed lines are the best-fit trends and legends show the Spearman rank correlation coefficients, which are also reported in Table~\ref{tab:1}.
    Each inset upper panel shows the rms distribution of the data after subtracting the best-fit trend in bins of the horizontal-axis variable. 
    }
    \label{fig:scatter-plot}
\end{figure*}

\begin{table*}[htb]
    \hspace{-1.5cm}
    \begin{tabular}{l c c c c c@{\hskip 2em} c c c c c}
    \hline\hline
    & \multicolumn{5}{c}{$\log_{10} I_{[\mathrm{C}\tt{I}](1\text{--}0)}/I_{\mathrm{CO}(2\text{--}1)}$} & \multicolumn{5}{c}{$\log_{10} I_{\mathrm{CO}(4\text{--}3)}/I_{\mathrm{CO}(2\text{--}1)}$} \\
    \cline{2-6} \cline{7-11}
    $x$ variable & $N$ & $A$ & $\rho$ & $P_0$ & rms & $N$ & $A$ & $\rho$ & $P_0$ & rms \\
    \hline
1) $\log_{10} I_{\mathrm{CO}(1\text{--}0)}$ & $0.38$ & $-1.94$ & $0.65$ & $<0.001$ & $0.16$ & $0.61$ & $-2.13$ & $0.74$ & $<0.001$ & $0.25$ \\
2) $\sigma_{\mathrm{CO}(2\text{--}1), \; 30 \, \mathrm{pc}}$ & $0.00$ & $-1.00$ & $0.48$ & $<0.001$ & $0.19$ & $0.01$ & $-0.65$ & $0.60$ & $<0.001$ & $0.28$ \\
3) $\log_{10} (I_{\mathrm{CO}(2\text{--}1)}/I_{\mathrm{CO}(1\text{--}0)})$ & $1.38$ & $-0.82$ & $0.37$ & $0.001$ & $0.18$ & $0.50$ & $-0.42$ & $0.21$ & $0.064$ & $0.30$ \\
4) $\log_{10} I_{\mathrm{2\mu\mathrm{m}}}$ & $0.25$ & $-1.29$ & $0.79$ & $<0.001$ & $0.15$ & $0.20$ & $-0.75$ & $0.67$ & $<0.001$ & $0.27$ \\
5) $\log_{10} I_{\mathrm{11.3\mu\mathrm{m}}}$ & $0.26$ & $-1.56$ & $0.84$ & $<0.001$ & $0.12$ & $0.16$ & $-0.83$ & $0.67$ & $<0.001$ & $0.26$ \\
6) $\log_{10} I_{\mathrm{21\mu\mathrm{m}}}$ & $0.18$ & $-1.41$ & $0.83$ & $<0.001$ & $0.13$ & $0.13$ & $-0.78$ & $0.70$ & $<0.001$ & $0.25$ \\
    \hline\hline\\
    \end{tabular}
    \caption{%
        Best-fit slope $N$ and intercept $A$ of the $y = N \cdot x + A$ lines shown in Fig.~\ref{fig:scatter-plot}, where $y$ is one of the $\log_{10} \Remark{RCICO} \equiv \log_{10} (I_{\mathrm{[C{\tt{I}}](1\text{--}0)}} / I_{\mathrm{CO(2\text{--}1)}})$ and $\log_{10} \Remark{R42} \equiv \log_{10} (I_{\mathrm{CO(4\text{--}3)}} / I_{\mathrm{CO(2\text{--}1)}})$ line ratios, and $x$ is one of the variables to correlate with: 
        \textit{1)} the CO(1--0) integrated intensity $\log_{10} I_{\mathrm{CO(1\text{--}0)}}$; 
        \textit{2)} the high-resolution ($\sim 30$~pc) CO(2--1) moment-2 map, convolved to the $3.5''$ ($\sim 330$~pc) resolution $\sigma_{\mathrm{CO(2\text{--}1)},\;30\,\mathrm{pc}}$; 
        \textit{3)} the low-$J$ CO line ratio $\log_{10} (I_{\mathrm{CO(2\text{--}1)}} / I_{\mathrm{CO(1\text{--}0)}})$; 
        \textit{4)} the stellar continuum dominated 2~\um{} emission $\log_{10} I_{\mathrm{2\mu\mathrm{m}}}$; 
        \textit{5)} the PAH dominated 11.3~\um{} emission $\log_{10} I_{\mathrm{11.3\mu\mathrm{m}}}$; 
        and \textit{6)} the dust continuum dominated 21~\um{} emission $\log_{10} I_{\mathrm{21\mu\mathrm{m}}}$. 
        For each of $\Remark{RCICO}$ and $\Remark{R42}$, we list five fitting result parameters: the best-fit slope $N$, the intercept $A$, the Spearman correlation coefficient $\rho$, the null hypothesis probability $P_0$, and the rms scatter of data points after subtracting the best-fit line. See Sect.~\ref{subsec:correlations} for details. 
    }
    \label{tab:1}
\end{table*}
%\end{center}

\subsection{Correlations of CO and \CI\ line ratios}
\label{subsec:correlations}

We investigate what star formation and ISM properties correlate with the CO and \CI{} line ratios in Fig.~\ref{fig:scatter-plot}. 
We examine CO(1--0) line-integrated intensity ($\Remark{ICO10}$), molecular gas velocity dispersion ($\sigma_{\mathrm{CO(2\text{--}1),\,30~pc}}$\,\footnote{For the molecular gas velocity dispersion, we use the native resolution CO(2--1) equivalent width (ew) map \citep{SCHINNERER_PHANGSJWST} then 
compute the average in apertures with size equaling the common resolution, so as to trace the velocity dispersion at $\sim 30$~pc.}), 
\Remark{R21}, 21~$\um$, PAH-dominated 11.3~$\um$, and stellar continuum dominated 2~$\um$ intensities, all at PSF-matched $3.5''$ ($\sim 330$~pc) resolution and sampled in independent resolution units (by binning in hexagons with diameter equaling the beam FWHM $3.5''$ then taking the aperture center value). 
For each pair of variables, we perform a linear fitting in log-log space with the \incode{scipy.optimize.curve_fit} code, including errors on the line ratio propagated the  original moment and uncertainty maps (Sect.~\ref{subsec:alma}). 
We obtain a best-fit slope $N$ and intercept $A$ for each fit and calculate the rms of the data points around the best-fitting line. We also calculate the Spearman correlation coefficient, $\rho$, and null hypothesis probability, $P_0$, using the \incode{pingouin} code. 
The resulting parameters are summarized in Table~\ref{tab:1}. 

The properties that correlate the most strongly with \Remark{R42} and \Remark{RCICO} are the PAH-dominated 11.3~$\um$, the dust continuum dominated 21~$\um$ and the stellar dominated 2~$\um$ emission, with $\rho > 0.7$. 
The $\Remark{ICO10}$ is the next strongly correlated variable. 
The velocity dispersion, $\sigma_{\mathrm{CO(2\text{--}1),\,30~pc}}$, and $\Remark{R21}$ show weak or no correlation, with $\rho \lesssim 0.3$. 

For each scatter plot panel in Fig.~\ref{fig:scatter-plot}, we show the rms distribution of the data in bins of the horizontal-axis variable. 
In all cases (except for $\Remark{R21}$, panel $c$) the rms is higher at lower $x$-axis values. 
For all variables except $\Remark{ICO10}$ (panel $a$), \Remark{R42} and \Remark{RCICO} tend to lie below the fitted line at the faint end.
In regions with low CO intensities, the CO(4--3) line can still be as highly excited as CO-brightest regions, despite a factor of $\sim 10$ weaker CO(1--0) emission. This may be surprising if the CO(1--0) emission indeed indicates lower gas surface density and perhaps volume density, but these regions might also simply show low CO(1--0) at 330~pc resolution because of a low area filling fraction. 
Obtaining higher-resolution high-$J$ CO and \CI\ observations is critical to understanding the scenarios in the future.

In panel $b)$ of Fig.~\ref{fig:scatter-plot}, we find no correlations of $\Remark{R42}$ or $\Remark{RCICO}$ versus $\sigma_{\mathrm{CO}}$, with slope $\sim 0$. This indicates that CO excitation and \CI\ enrichment correlate more strongly with radiation properties than with gas dynamics in this starburst ring. 
Indeed, in Fig.~\ref{fig:1} (g), we find a peak $\sigma_{\mathrm{CO}} \sim 50 \; \mathrm{km\,s^{-1}}$ at the Southern Arm, whereas the Northern Arm with the strongest star formation and most numerous YMCs has nearly half that value, i.e., $\sigma_{\mathrm{CO}} \sim 20\text{--}30 \; \mathrm{km\,s^{-1}}$. 
\cite{SCHINNERER_PHANGSJWST} find that the CO emission in this system is usually composed of multiple (e.g. 2--4) line-of-sight velocity components, with individual component having a similar line width $\sigma \sim 19 \; \mathrm{km \, s^{-1}}$. Therefore, this geometry effect may well blend out any $\sigma_{\mathrm{CO}}$ versus line ratios trends in this study.

In panel $c)$ of Fig.~\ref{fig:scatter-plot}, the poor correlation between \Remark{R21} and \Remark{R42} at our $\sim 330$~pc resolution indicates that the low-$J$ and mid-$J$ CO SLED shapes are largely decoupled. $\Remark{R21}$ is mostly saturated/thermalized in environments like the center of NGC~1365. It is therefore necessary to obtain higher-$J$ CO lines to trace the CO excitation.

The stellar dominated 2~\um, PAH-dominated 11.3~\um, and warm dust continuum dominated 21~\um\ emission all show tight, positive correlations with the line ratios as seen in panels $d)$ to $f)$ of Fig.~\ref{fig:scatter-plot}. These correlations are robust (i.e., tighter than about 0.2~dex) in the range of 
$10 \lesssim I_{2\,\um} / [\mathrm{MJy\,sr^{-1}}] \lesssim 80$, 
$20 \lesssim I_{11.3\,\um} / [\mathrm{MJy\,sr^{-1}}] \lesssim 1000$, and
$20 \lesssim I_{21\,\um} / [\mathrm{MJy\,sr^{-1}}] \lesssim 2000$, respectively. 
The 2~\um\ correlation has the smallest valid range, only less than a decade, whereas the dust correlations are valid over nearly two decades. 
The galaxy center and its surrounding apertures are outliers in the line ratio versus $I_{2\,\um}$ plot even at our $\sim330$~pc resolution, but not for the line ratios versus PAH or dust continuum. 
We caution that the statistics is based on only one starburst ring system at a coarse resolution. 
Larger-sample studies will be critical to deeper understanding of the statistics.

Comparing the trends in \Remark{R42} to those in \Remark{RCICO}, we find similar correlations relating each line ratio to the other variables. The trends with \Remark{RCICO} do tend to be tighter, with scatter about $\sim 0.05$--0.1~dex smaller than we find for \Remark{R42}. 
This likely relates to the fact that $\Remark{CI10}$ has a $>30\times$ lower critical density and $2 \times$ lower upper level energy temperature ($E_u$) than CO(4--3) \citep[e.g.,][Table~3]{Crocker2019}. This makes $\Remark{CI10}$ more sensitive to the temperature and low-density part of the medium, which is substantially affected by the stellar feedback from embedded star formation traced by the YMCs, PAH emission, and warm dust continuum.

\section{Summary}

In this Letter we use multi-$J$ CO and \CI\ line ratios to trace the CO excitation and dissociation and infer molecular gas temperature, density and feedback under the impact of YMCs in the bar-fed starbursting ring of NGC~1365. 
The mid-$J$ CO SLED up to CO(4--3), and the $\Remark{CI10}$ line, together with the distribution of young ($< 10 \ \mathrm{Myr}$), massive ($M_{\star} \gtrsim 10^{6} \ \mathrm{M_{\odot}}$) star clusters revealed by JWST
allow us to infer 
how the molecular gas properties are impacted by stellar feedback from YMCs as the gas enters and circulates in the starburst ring.
We summarize our findings below.

\begin{itemize}
    
    \item The Northern and Southern Arms have a high molecular gas density, relatively low temperature and $\XCICO$ abundance ratio, with observed line ratios $\Remark{R21} \sim 0.8 \text{--} 1$, $\Remark{R42} \sim 0.45 \text{--} 0.51$ and $\Remark{RCICO} \sim 0.1 \text{--} 0.2$. These are in line with the scenario where the Northern and Southern Arm regions are the locations where molecular gas flows into the starburst ring along the bar and dust lanes. Bar-induced shocks may play a key role in affecting the gas there, but further observational support is needed. 
    
    \item The molecular gas in the Mid-Southwest region seems largely impacted by the stellar feedback (i.e., low $\Remark{R42}$ but high $\Remark{R21}$ and $\Remark{RCICO}$), exhibiting a low gas density ($\log (\nHtwo/\mathrm{cm^{-3}}) \sim 2.9$), high temperature ($\sim 100$~K) and enhanced $\XCICO$ abundance ratio ($\gtrsim 1$) compared to the upstream Northern Arm ($\log (\nHtwo/\mathrm{cm^{-3}}) \sim 3.4$, $\Tkin \sim 40$~K and $\XCICO \sim 0.7$) as inferred from our LVG fitting. 
    
    \item The molecular gas in the Mid-East region 
    exhibits both a high $\Remark{R42}$ and $\Remark{RCICO}$, possibly due to its much higher density than that of the Mid-Southwest region and thus less impacted by the stellar feedback. 
    Our LVG fitting infer that there is only a moderate decrease in gas density ($\log (\nHtwo/\mathrm{cm^{-3}}) \sim 3.4$) or weak increase in temperature ($\sim 25$~K) and $\XCICO$ ($\sim 0.5$) compared to the upstream Southern Arm region ($\log (\nHtwo/\mathrm{cm^{-3}}) \sim 4.0$, $\Tkin \sim 15$~K and $\XCICO \sim 0.2$). 

    \item Through a correlation analysis, we find that the mid-$J$ CO excitation $\Remark{R42}$ or the CO dissociation tracer $\Remark{RCICO}$ does not obviously correlate with the low-$J$ CO excitation $\Remark{R21}$, likely because the $\Remark{R21}$ shows high ratios and appears nearly thermalized across the whole region. 
    We also find little correlation between the line ratios and the apparent CO line velocity dispersion, 
    implying that the complex gas dynamics does not affect the CO excitation and photodissociation in the starburst ring.
    
    \item We find tightest correlations between $\Remark{R42}$ or $\Remark{RCICO}$ and the mid-IR PAH-dominated 11.3~\um{} and dust continuum dominated 21~\um{} emission 
    ($\rho \sim 0.67$--0.84, $\mathrm{rms} \sim 0.12$--0.27~dex). The stellar continuum dominated 2~\um{} emission correlates with $\Remark{R42}$ and $\Remark{RCICO}$ well too 
    ($\rho \sim 0.67$--0.79, $\mathrm{rms} \sim 0.15$--0.27~dex) but the very center does not follow the trend.
    
    \item The $\Remark{RCICO}$ correlations with mid-IR dust/PAH and near-IR stellar emission properties are in general slightly tighter ($\sim 0.1$ higher in $\rho$) and less scattered ($\sim -0.5$~dex smaller rms) than those of $\Remark{R42}$. This may relate to the significantly lower critical density of $\Remark{CI10}$ than $\Remark{CO43}$ and to CO dissociation which make \CI{} more sensitive to the mid-IR traced bulk of star-forming gas and stellar feedback.
    
    \item Despite hosting an Seyfert 1.5 AGN and having an ionized gas outflow, NGC~1365's central $\sim 330$~pc area (our resolution unit) exhibits only moderate CO excitation and \CI/CO line ratio comparable to or even less highly excited than other regions that we studied. 
\end{itemize}

\vspace{2cm}

\appendix

\counterwithin{figure}{section}

\section{LVG model fitting}
\label{appendix:LVG}

\begin{figure}[htb]
\centering
\includegraphics[width=0.49\linewidth]{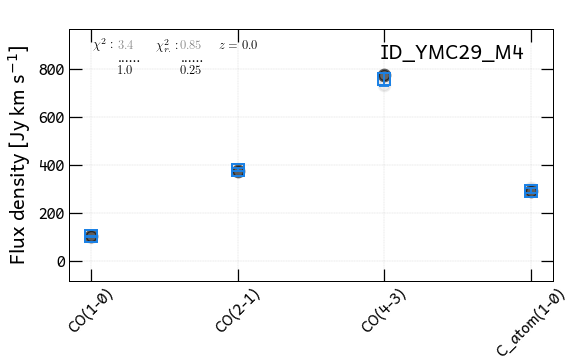}
\includegraphics[width=0.49\linewidth]{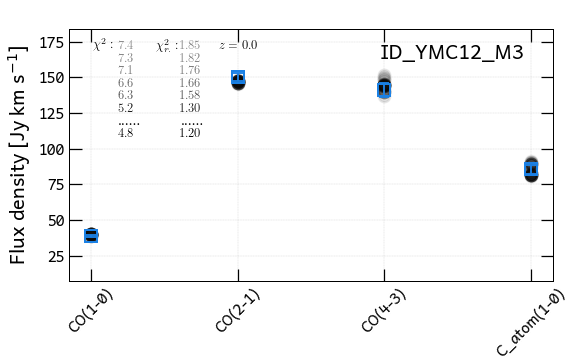}\\
\includegraphics[width=0.49\linewidth]{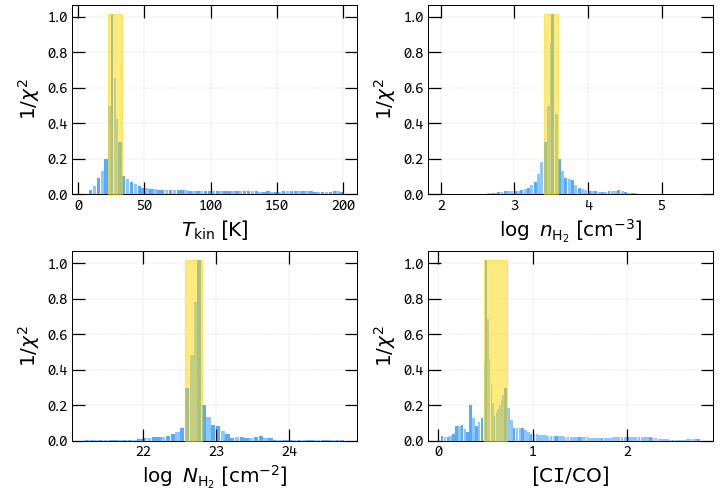}
\includegraphics[width=0.49\linewidth]{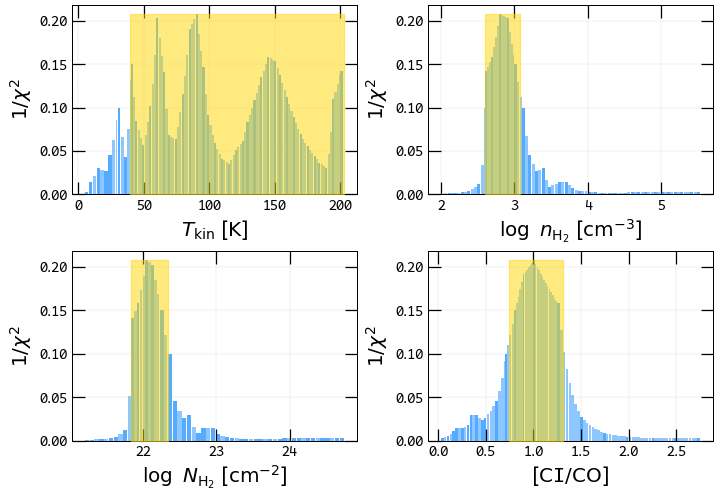}
\vspace{0.5ex}
\caption{%
\textit{Upper row}: the CO and \CI\ SLED of YMC~29 (M4, the most excited YMC aperture in the Northern Arm; \textit{left} panel) and that of YMC~12 (M3, the least excited YMC aperture in the Mid-Southwest region; \textit{right} panel). 
Observed line fluxes are shown as blue squares with error bars. 
Black to gray solid circles are fitted points from a series of fitting within $1\sigma$ with an increasing $\chi^2$ (and reduced $\chi^2$, $\chi_{r}^2$). 
\textit{Bottom panels}: corresponding $1/\chi^2$ distributions of model fitting parameters $\Tkin$, $\lgnHtwo$, $\log N_{\mathrm{H_2}}$ (non-independent) and $\XCICO$ from the \incode{MICHI2} code. 
Yellow shading indicates the $\pm 1\sigma$ parameter range.
\label{fig:LVG fitting}
}
\end{figure}

We illustrate our Monte Carlo LVG model fitting in Fig.~\ref{fig:LVG fitting}. We first use RADEX \citep{vanderTak2007} with the Leiden Atomic and Molecular Database \citep{Schoier2005} to build a library of LVG models with the parameter grids as described in Sect.~\ref{subsec:zoom-in}. 
Then, we use the \incode{MICHI2} code \citep{Liudz2021} to run Monte Carlo fitting and obtain the $1/\chi^2$ posterior distribution for each parameter (following the statistical criterion of $\pm 1\sigma$ in \citealt{Press1992}). 
Blue squares with error bars are the CO and \CI\ line fluxes to be fitted, and black to gray dots are model data points with different reduced $\chi^2$. 
Our free model parameters are $\lgnHtwo$, $\Tkin$, $\XCICO$, and normalization (beam filling factor). 
Their $1/\chi^2$ distributions and the $\pm 1\sigma$ ranges are shown in the lower panels. 
All fitting results are presented in Table~\ref{tab:2}.

\begin{table}[htb]
\begin{center}
\begin{tabular*}{0.89\linewidth}{lcccccc}
\hline\hline
Aperture & RA (J2000) & DEC (J2000) & $\log n_{\mathrm{H_2}} / \mathrm{cm^{-3}}$ & $\log N_{\mathrm{H_2}} / \mathrm{cm^{-2}}$ & $T_{\mathrm{kin}} / \mathrm{K}$ & $[\mathrm{C}\text{\sc{i}}/\mathrm{CO}]$ \\
\hline
\textit{Northern Arm} &  &  &  &  &  &  \\
CO.N2 & $53.40340917$ & $-36.13729972$ & $3.20_{-0.12}^{+0.25}$ & $22.44_{-0.12}^{+0.26}$ & $25.0_{-11.2}^{+8.0}$ & $0.60_{-0.25}^{+0.13}$ \\
36 & $53.40311543$ & $-36.13715197$ & $2.70_{-0.02}^{+0.35}$ & $21.94_{-0.01}^{+0.36}$ & $155.0_{-110.0}^{+12.9}$ & $1.00_{-0.31}^{+0.09}$ \\
35 & $53.40306629$ & $-36.13749402$ & $3.30_{-0.26}^{+0.09}$ & $22.54_{-0.25}^{+0.06}$ & $25.0_{-2.3}^{+24.7}$ & $0.60_{-0.03}^{+0.40}$ \\
33,M5 & $53.40251420$ & $-36.13768031$ & $3.30_{-0.13}^{+0.09}$ & $22.54_{-0.16}^{+0.06}$ & $35.0_{-3.4}^{+13.0}$ & $0.80_{-0.08}^{+0.16}$ \\
30 & $53.40098220$ & $-36.13841668$ & $3.30_{-0.17}^{+0.09}$ & $22.54_{-0.16}^{+0.06}$ & $35.0_{-3.4}^{+18.0}$ & $0.70_{-0.04}^{+0.29}$ \\
29,M4 & $53.40170236$ & $-36.13843398$ & $3.50_{-0.11}^{+0.09}$ & $22.74_{-0.16}^{+0.07}$ & $25.0_{-2.3}^{+8.0}$ & $0.50_{-0.02}^{+0.23}$ \\
28,M6 & $53.40318107$ & $-36.13847524$ & $3.60_{-0.25}^{+0.09}$ & $22.84_{-0.20}^{+0.07}$ & $15.0_{-1.2}^{+12.9}$ & $0.30_{-0.03}^{+0.36}$ \\
25 & $53.40092801$ & $-36.13910029$ & $3.40_{-0.08}^{+0.10}$ & $22.64_{-0.09}^{+0.04}$ & $20.0_{-3.1}^{+2.9}$ & $0.40_{-0.04}^{+0.02}$ \\
CO.N1 & $53.40132208$ & $-36.13871556$ & $3.40_{-0.05}^{+0.14}$ & $22.64_{-0.06}^{+0.17}$ & $30.0_{-10.2}^{+3.0}$ & $0.60_{-0.20}^{+0.06}$ \\
\textit{Mid-Southwest Region} &  &  &  &  &  &  \\
22 & $53.40046102$ & $-36.13984982$ & $3.30_{-0.03}^{+0.02}$ & $22.54_{-0.06}^{+0.01}$ & $20.0_{-3.1}^{+2.9}$ & $0.45_{-0.03}^{+0.03}$ \\
19 & $53.39952417$ & $-36.14009082$ & $2.60_{-0.01}^{+0.85}$ & $21.84_{+0.00}^{+0.87}$ & $195.0_{-180.0}^{+7.9}$ & $1.00_{-0.70}^{+0.09}$ \\
12,M3 & $53.39955255$ & $-36.14112108$ & $2.80_{-0.20}^{+0.27}$ & $22.04_{-0.20}^{+0.30}$ & $90.0_{-50.6}^{+112.9}$ & $1.00_{-0.25}^{+0.31}$ \\
11 & $53.39977124$ & $-36.14112797$ & $2.70_{-0.02}^{+0.35}$ & $21.94_{-0.01}^{+0.36}$ & $175.0_{-125.8}^{+17.9}$ & $1.25_{-0.35}^{+0.12}$ \\
9 & $53.40018429$ & $-36.14131523$ & $2.70_{-0.02}^{+0.39}$ & $21.94_{-0.08}^{+0.36}$ & $180.0_{-130.8}^{+17.9}$ & $1.25_{-0.41}^{+0.15}$ \\
7,M2 & $53.39987072$ & $-36.14183303$ & $3.10_{-0.42}^{+0.09}$ & $22.34_{-0.41}^{+0.06}$ & $35.0_{-3.4}^{+157.9}$ & $0.70_{-0.07}^{+0.58}$ \\
5 & $53.40033806$ & $-36.14207205$ & $3.30_{-0.26}^{+0.04}$ & $22.54_{-0.30}^{+0.06}$ & $20.0_{-0.2}^{+17.4}$ & $0.40_{-0.01}^{+0.30}$ \\
3,J & $53.40069177$ & $-36.14212741$ & $3.40_{-0.08}^{+0.05}$ & $22.64_{-0.09}^{+0.04}$ & $15.0_{-4.0}^{+4.9}$ & $0.30_{-0.03}^{+0.03}$ \\
\textit{Mid-East Region} &  &  &  &  &  &  \\
23 & $53.40337423$ & $-36.13956883$ & $3.30_{-0.04}^{+0.35}$ & $22.54_{-0.03}^{+0.36}$ & $30.0_{-16.1}^{+3.0}$ & $0.60_{-0.36}^{+0.06}$ \\
21 & $53.40333174$ & $-36.13986047$ & $3.40_{-0.10}^{+0.14}$ & $22.64_{-0.06}^{+0.17}$ & $25.0_{-5.2}^{+2.9}$ & $0.45_{-0.12}^{+0.14}$ \\
18 & $53.40339461$ & $-36.14021144$ & $3.60_{-0.25}^{+0.05}$ & $22.84_{-0.20}^{+0.07}$ & $15.0_{-1.2}^{+12.9}$ & $0.25_{-0.03}^{+0.23}$ \\
16 & $53.40291154$ & $-36.14045529$ & $3.40_{-0.05}^{+0.14}$ & $22.64_{-0.06}^{+0.17}$ & $25.0_{-5.2}^{+4.7}$ & $0.45_{-0.15}^{+0.05}$ \\
15 & $53.40321229$ & $-36.14061030$ & $3.50_{-0.06}^{+0.14}$ & $22.74_{-0.05}^{+0.16}$ & $20.0_{-5.0}^{+4.7}$ & $0.30_{-0.10}^{+0.05}$ \\
13 & $53.40293504$ & $-36.14082502$ & $3.50_{-0.11}^{+0.09}$ & $22.74_{-0.10}^{+0.07}$ & $20.0_{-3.2}^{+7.9}$ & $0.30_{+0.00}^{+0.18}$ \\
10 & $53.40283403$ & $-36.14127466$ & $3.60_{-0.23}^{+0.05}$ & $22.84_{-0.25}^{+0.07}$ & $15.0_{-1.2}^{+14.7}$ & $0.20_{+0.00}^{+0.30}$ \\
\textit{Southern Arm} &  &  &  &  &  &  \\
1 & $53.40028097$ & $-36.14257775$ & $3.50_{-0.05}^{+0.05}$ & $22.74_{-0.06}^{+0.00}$ & $15.0_{-1.1}^{+4.9}$ & $0.25_{-0.01}^{+0.02}$ \\
2 & $53.39989135$ & $-36.14257526$ & $3.50_{-0.11}^{+0.04}$ & $22.74_{-0.10}^{+0.07}$ & $15.0_{-1.2}^{+7.9}$ & $0.25_{-0.01}^{+0.13}$ \\
CO.S1 & $53.39917000$ & $-36.14308806$ & $4.10_{-0.01}^{+0.04}$ & $23.34_{-0.01}^{+0.06}$ & $10.0_{-2.0}^{+3.9}$ & $0.05_{-0.02}^{+0.01}$ \\
CO.S2 & $53.39845708$ & $-36.14325417$ & $4.00_{-0.05}^{+0.05}$ & $23.24_{-0.04}^{+0.03}$ & $10.0_{-2.0}^{+1.0}$ & $0.05_{-0.02}^{+0.01}$ \\
CO.S3 & $53.39761333$ & $-36.14309889$ & $3.50_{-0.09}^{+0.05}$ & $22.74_{-0.06}^{+0.00}$ & $15.0_{-1.1}^{+4.9}$ & $0.15_{-0.03}^{+0.03}$ \\
\textit{Center} &  &  &  &  &  &  \\
Center & $53.40154167$ & $-36.14041694$ & $3.50_{-0.24}^{+0.04}$ & $22.74_{-0.23}^{+0.07}$ & $15.0_{-1.2}^{+12.9}$ & $0.25_{-0.01}^{+0.23}$ \\
\hline\hline
\end{tabular*}
\end{center}
\caption{%
Results of \incode{MICHI2} LVG fitting to the CO and \CI\ SLED of YMC, CO peak and galaxy center apertures. 
YMC~IDs and R.A. and Declination coordinates are from \cite{WHITMORE_PHANGSJWST}. See description of the fitting in Sect.~\ref{subsec:zoom-in}. See also Fig.~\ref{fig:LVG} for the illustration of the fitted parameters along the starburst ring. 
\label{tab:2}
}
\end{table}

\section*{Acknowledgments}

We thank the anonymous referee for very helpful comments. 
This work was carried out as part of the PHANGS collaboration.
This work is based on observations made with the NASA/ESA/CSA JWST and NASA/ESA Hubble Space Telescopes. The data were obtained from the Mikulski Archive for Space Telescopes at the Space Telescope Science Institute, which is operated by the Association of Universities for Research in Astronomy, Inc., under NASA contract NAS 5-03127 for JWST and NASA contract NAS 5-26555 for HST. The JWST observations are associated with program 2107, and those from HST with program 15454.

Some of the data presented in this paper were obtained from the Mikulski Archive for Space Telescopes (MAST) at the Space Telescope Science Institute. 
The specific observations analyzed can be accessed via:
\dataset[PHANGS-JWST observations]{http://dx.doi.org/10.17909/9bdf-jn24},
\dataset[PHANGS-HST image products]{https://doi.org/10.17909/t9-r08f-dq31} 
and
\dataset[PHANGS-HST catalog products]{https://doi.org/10.17909/jray-9798}. %(\url{https://archive.stsci.edu/hlsp/phangs-hst}). 

This Letter makes use of the following ALMA data: 
\incode{ADS/JAO.ALMA#2019.1.01635.S}, 
\incode{ADS/JAO.ALMA#2013.1.01161.S},
\incode{ADS/JAO.ALMA#2015.1.01135.S},
\incode{ADS/JAO.ALMA#2017.1.00129.S}.
ALMA is a partnership of ESO (representing its member states), NSF (USA) and NINS (Japan), together with NRC (Canada), MOST and ASIAA (Taiwan), and KASI (Republic of Korea), in cooperation with the Republic of Chile. The Joint ALMA Observatory is operated by ESO, AUI/NRAO and NAOJ.

AKL gratefully acknowledges support by grants 1653300 and 2205628 from the National Science Foundation, by award JWST-GO-02107.009-A, and by a Humboldt Research Award from the Alexander von Humboldt Foundation.
AU acknowledges support from the Spanish grants PGC2018-094671-B-I00, funded by MCIN/AEI/10.13039/501100011033 and by ``ERDF A way of making Europe'', and PID2019-108765GB-I00, funded by MCIN/AEI/10.13039/501100011033. 
ER acknowledges the support of the Natural Sciences and Engineering Research Council of Canada (NSERC), funding reference number RGPIN-2022-03499.
JMDK gratefully acknowledges funding from the European Research Council (ERC) under the European Union's Horizon 2020 research and innovation programme via the ERC Starting Grant MUSTANG (grant agreement number 714907). COOL Research DAO is a Decentralized Autonomous Organization supporting research in astrophysics aimed at uncovering our cosmic origins.
MC gratefully acknowledges funding from the DFG through an Emmy Noether Research Group (grant number CH2137/1-1).
SCOG, RSK, EJW acknowledge funding provided by the Deutsche Forschungsgemeinschaft (DFG, German Research Foundation) -- Project-ID 138713538 -- SFB 881 (``The Milky Way System'', subprojects A1, B1, B2, B8, and P2). 
JS acknowledges support by the Natural Sciences and Engineering Research Council of Canada (NSERC) through a Canadian Institute for Theoretical Astrophysics (CITA) National Fellowship.
FB and JdB would like to acknowledge funding from the European Research Council (ERC) under the European Union’s Horizon 2020 research and innovation programme (grant agreement No.726384/Empire).
KG is supported by the Australian Research Council through the Discovery Early Career Researcher Award (DECRA) Fellowship DE220100766 funded by the Australian Government. 
KG is supported by the Australian Research Council Centre of Excellence for All Sky Astrophysics in 3 Dimensions (ASTRO~3D), through project number CE170100013. 
HAP acknowledges support by the National Science and Technology Council of Taiwan under grant 110-2112-M-032-020-MY3.
RSK acknowledges financial support from the European Research Council via the ERC Synergy Grant ``ECOGAL'' (project ID 855130), from the Heidelberg Cluster of Excellence (EXC 2181 - 390900948) ``STRUCTURES'', funded by the German Excellence Strategy, and from the German Ministry for Economic Affairs and Climate Action in project ``MAINN'' (funding ID 50OO2206). 
SD is supported by funding from the European Research Council (ERC) under the European Union’s Horizon 2020 research and innovation programme (grant agreement no. 101018897 CosmicExplorer).
OE gratefully acknowledge funding from the Deutsche Forschungsgemeinschaft (DFG, German Research Foundation) in the form of an Emmy Noether Research Group (grant number KR4598/2-1, PI Kreckel). 
JP acknowledges support by the DAOISM grant ANR-21-CE31-0010 and by the Programme National ``Physique et Chimie du Milieu Interstellaire'' (PCMI) of CNRS/INSU with INC/INP, co-funded by CEA and CNES.
TGW acknowledges funding from the European Research Council (ERC) under the European Union’s Horizon 2020 research and innovation programme (grant agreement No. 694343).
SKS acknowledges financial support from the German Research Foundation (DFG) via Sino-German research grant SCHI 536/11-1.

\software{%
Astropy \citep{astropy:2013, astropy:2018, astropy:2022},
CASA software \citep{CASA}, 
Matplotlib \citep{matplotlib},
MICHI2 \citep{Liudz2021,michi2},
PHANGS-ALMA pipeline \citep{Leroy2021PhangsAlmaPipeline}, 
Photutils \citep{photutils}, 
RADEX \citep{vanderTak2007}, 
Scipy \citep{scipy},
spectral\_cube \citep{spectralcube}
}

%\bibliography{Biblio,phangsjwst}

\bibliographystyle{aasjournal}

\suppressAffiliationsfalse
\allauthors

\end{document}